\documentclass[fleqn,usenatbib]{mnras}
\usepackage{newtxtext,newtxmath}

\usepackage[T1]{fontenc}
\usepackage{ae,aecompl}

\usepackage{mathrsfs}

\usepackage{graphicx}	
\usepackage{multirow}

\newcommand{\HS}{HS~1700+6416} 
\newcommand{\HE}{HE~2347-4342} 
\newcommand{\HST}{{\itshape HST}} 
\newcommand{\FUSE}{{\itshape FUSE}} 
\newcommand{\VLT}{{\itshape VLT}} 
\newcommand{\Keck}{{\itshape Keck}} 
\newcommand{\Secref}[1]{\hyperref[#1]{Section~\ref{#1}}}
\newcommand{\subSecref}[1]{\hyperref[#1]{Subsection~\ref{#1}}}

\newcommand{\ovi}{\ion{O}{VI}}
\newcommand{\civ}{\ion{C}{IV}}
\newcommand{\hi}{\ion{H}{I}}
\newcommand{\heii}{\ion{He}{II}}

\newcommand{\taulya}{\tau_{\hi~Ly\mbox{-}\alpha}}
\newcommand{\lya}{Ly-$\alpha$}
\newcommand{\GP}{GP}

\title[Probing Inhomogeneity in the UV Background]{Probing Inhomogeneity in the Helium Ionizing UV Background}

\author[Morrison et al.]{
Sean Morrison,$^{1}$\thanks{E-mail: sean.morrison@lam.fr}
Matthew M. Pieri,$^{1}$
David Syphers,$^{2}$
and Tae-Sun Kim$^{3}$
\\
$^{1}$Aix Marseille Univ, CNRS, CNES, LAM, Marseille, France\\
$^{2}$Eastern Washington University, Cheney, WA\\
$^{3}$University of Wisconsin Madison
}

\date{Accepted XXX. Received YYY; in original form ZZZ}

\pubyear{2019}

\begin{document}
\label{firstpage}
\pagerange{\pageref{firstpage}--\pageref{lastpage}}
\maketitle

\begin{abstract}
We present an analysis combining the simultaneous measurement of intergalactic absorption by hydrogen (\hi), helium (\heii) and oxygen (\ovi) in UV and optical quasar spectra.
The combination of the \hi\ and \heii\ Lyman-alpha forests through $\eta$ (the ratio of column densities of singly ionized helium to neutral hydrogen) is thought to be sensitive to large-scale inhomogeneities in the extragalactic UV background.
We test this assertion by measuring associated five-times-ionized oxygen (\ovi) absorption, which is also sensitive to the UV background. We apply the pixel optical depth technique to \ovi\ absorption in high and low $\eta$ samples filtered on various scales. This filtering scale is intended to represent the dominant scale of any coherent oxygen excess/deficit. We find a $2\sigma$ detection of an \ovi\ opacity excess in the low $\eta$ sample on scales of $\sim$10~cMpc for \HE\ at $\bar{z}\approx 2.6$, consistent with a large-scale excess in hard UV photons. However, for \HS\ at $\bar{z}\approx 2.5$ we find that the measured \ovi\ absorption is not sensitive to differences in $\eta$.
\HS\ also shows a relative absence of \ovi\ overall, which is $6\sigma$ inconsistent with that of \HE. This implies UV background inhomogeneities on $\gtrsim$200~cMpc scales, hard UV regions having internal ionization structure on $\sim$10~cMpc scales and soft UV regions showing no such structure.
Furthermore, we perform the pixel optical depth search for oxygen on the \heii\ Gunn-Peterson trough of \HE\  and find results consistent with post-\heii-reionization conditions.

\end{abstract}



\section{Introduction}
After the formation of the cosmic microwave background, the universe recombined to produce a homogeneous, neutral, and diffuse medium. Between $6<z<30$ the universe transitioned from a predominately neutral to a predominately ionized state \citep{Fan2006}. In the period leading to the end of reionization, enough neutral hydrogen (\hi) remained to generate a Gunn-Peterson (GP) trough in quasar absorption, where-by blanket absorption is seen bluewards of the quasar Lyman-$\alpha$ ($\lambda 1216$; \lya) emission line. At lower redshifts this blanket is lifted and a \hi\ \lya\ forest is observed reflecting the residual neutral gas. Helium is singly ionized during this period, which (given its status as a hydrogen-like ion) gives rise to the helium-II \lya\ ($\lambda 304$) absorption. 

Following the epoch of hydrogen reionzation, quasars and other ionizing sources proceeded to further ionize singly-ionized helium (\heii) in a stage known as helium-II reionization. \cite{shull2010} claims that this epoch ended around 
$z_r\sim 2.7$, though other estimates claim it ended as early as $z_r\sim 3.2$ \citep{syphers2011,Worseck2016}. This is a subject of ongoing study. 

Various source populations contribute to the reionzation of the universe and keep it ionized until the present day. The dominant sources of ionizing UV photons are thought to be active galactic nuclei associated with quasars and star-forming galaxies. Quasars contribute proportionately more hard UV photons and so their growing contribution leads to the epoch of \heii\ reionization \citep{Furlanetto2008}. Regions of excess ionizing photons around individual sources are known as `proximity zones'. The extent of the excess depend on factors such as the photon escape fraction of the surrounding gas, as well as the amount of time that the source has been active.

Fluctuations in the UV background may give rise to spatial variation in \hi\ and \heii.
However, when the mean free path of ionizing photos is  larger than the mean separation of closest sources, such effects are suppressed and potentially insignificant. At $z\sim 2-3$ the mean free path of \hi\ ionizing photons is $\sim100-300~ h^{-1}$ comoving megaparsec (cMpc) \citep{Croft2004, Rudie2013}, which is larger than the mean separation of the sources. 
The mean free path of harder \heii\ ionizing photos is significantly shorter in this redshift range. Various factors play a role in its determination (e.g. \citealt{Furlanetto2009,Davies2014,Faucher-Giguere2009,Davies2017}) leading to scales of 20-100~cMpc at $2.5<z<3.2$. Most notably, the mean free path itself is thought to vary spatially \citep{Davies2017} contributing to UV background fluctuations on scales of up to $\sim$~200~cMpc.

In the fluctuating Gunn-Peterson approximation (FGPA), \hi\ \lya\ opacity traces gas overdensities \citep{Croft1998,Weinberg1998}. 
One justifying factor for this approximation is the large mean free path of \hi\ ionizing  photons above.
Using the FGPA, \hi\ opacity $\tau_\ion{H}{I}$ can be used as an approximation of the intergalactic medium (IGM) density, while \ion{He}{II} opacity $\tau_\ion{He}{II}$ probes both the UV background and the IGM density. We can therefore combine these two quantities into a largely density independent ratio 
\begin{equation}
\eta \equiv \frac{N_\ion{He}{II}}{N_\ion{H}{I}}\approx 4\frac{\tau_\ion{He}{II}}{\tau_\ion{H}{I}}
\label{eta}
\end{equation}
to act as the standard observable for the hardness of the UV Ionizing background. The factor of four arises from the ratio of wavelengths (ionization energies) between He and H. The approximation holds if line profiles are dominated by turbulent broadening over thermal broadening (since turbulent broadening affects species equally while thermal broadening does not).
Large $\eta$ equates to soft radiation field, while small $\eta$ equates to a hard radiation field.

The ratio $\eta$ is sensitive to various physical effects.
\citet{Graziani2019} used radiative transfer simulations to show that $\eta$ is sensitive to intergalactic medium opacities on Mpc scales. 
On the other hand, $\eta$ is also affected by systematic effects not directly related to ionization, the most dominant being the different thermal line broadening between the \hi\ and \heii\ \citep{Fechner2007_461,McQuinn2014}. Various studies have examined the use of $\eta$ as a probe of the UV background fluctuations. \citet{McQuinn2014} argued that $\eta$ shows no large fluctuations, and that fluctuations found in early studies were the result of continuum errors and low signal-to-noise. However, \citet{Syphers2014} found $\eta$ sensitivity to UV background fluctuations (using higher signal-to-noise data compared to early studies and more refined continuum estimates).

Since the main observables are quasars themselves, the proximity effect is often split into line-of-sight (where the source is the observed quasar) and transverse (where the line-of-sight passes through the sphere of influence of another quasar). Given that quasar luminosity is variable,
fossil proximity zones around inactive quasars must also be considered \citep{Oppenheimer2013,Oppenheimer2017}.
\cite{Segers2017} found that these proximity zones could maintain elevated densities of high ions for timescales much greater than AGN lifetimes.

\begin{table*}
\caption{Ionization energies for a sample of atomic species \citep{NIST_ASD}}
\label{tab:IonEng}
\begin{tabular}{ccccccccc}
\hline
\multirow{3}{*}{Element} & \multicolumn{7}{c}{Ionization Energy (ev)}\\
{} & I & II & III & IV & V & VI & VII & VIII\\
\hline
\hline
H	 & 13.59843449 & \ldots 		& \ldots		& \ldots		& \ldots		&\ldots		&\ldots		&\ldots		\\
\hline
He	 & 24.58738880 & 54.4177650 	& \ldots		& \ldots		& \ldots		&\ldots		&\ldots		&\ldots		\\
\hline
O	 & 13.618055 	& 35.12112 	& 54.93554	& 77.41350	& 113.8990	& 138.1189	& 739.32682	& 871.40988	\\
\hline
\end{tabular}
\end{table*}

In addition to utilising \hi\ and \heii\ in the form of $\eta$, additional information can be obtained from the ratios of optical depths for various ionization species \citep{Songaila1995,Bolton2011}. Since different species ionize when absorbing photons of different energies (\autoref{tab:IonEng}), it is therefore possible to examine the abundance of various species to determine the spectral shape and intensity of the ionizing UV background \citep{Agofonova2007}. 
Furthermore, we may use the spectral shape of the UV background, its intensity and the size, shape, and location of any inhomogeneities to study the ionizing sources. Attempts have been made to probe the UV background shape and \heii\ reionization with \ion{C}{IV} and \ion{Si}{IV} observationally \citep{Songaila1995} and using toy models \citep{Bolton2011}. 
On the other hand, five-times ionized oxygen (\ovi; $\lambda\lambda 1032,1038$) is a potentially useful as a tracer of hard UV conditions, and so can probe to the nature of the UV background when combined with $\eta$.

In this paper we present a new analysis combining measurements of $\eta$ with intergalactic oxygen absorption in spectra of quasars \HE\ and \HS. We combined the forward modelling approach for calculating $\eta$ with the pixel optical depth method for measuring \ovi\ \citep{Cowie1998}. We
treat $\eta$ as a good proxy for UV hardness and cut the \ovi\ sample based on this proxy, with a simple high/low $\eta$ split. We explore the filtering of $\eta$ to study various physical scales and take novel approaches to analysing differences in \ovi\ pixel optical depth samples. 
We compare the overall difference in \ovi\ seen between our pair of quasar spectra to study line-of-sight scales.
We further compare oxygen absorption in the \heii\ GP trough to the \heii\ forest in order to search for strong evolution associated with \heii\ reionization.

We study large-scale inhomogeneities in the UV background both within lines-of-sight and between lines-of-sight. By using $\eta$ as a proxy for excess hard UV photons rather than the location of observed quasars, we are able to including the contribution of fossil quasar proximity effects (assuming similar recombination times for \heii\ and \ovi). Despite our small analysis sample we are able to study inhomogeneous hardening of the UV background at and after \heii\ reionization.

This paper is structured as follows:
\Secref{sec:Data} describes the data and data reduction techniques.
\Secref{sec:etaForwardModeling} outlines the forward modeling used to determine $\eta$.
\Secref{sec:POD} gives a brief overview of the pixel optical depth method (POD) and then details our use of it in combination with high-low splits in $\eta$.
\Secref{sec:dT} shows our results calculating differential oxygen opacity for various pixel sample splits. First we show the separation into high-low splits in $\eta$ filtered on various physical scales, then the difference between lines-of-sight, and finally splitting the sample between GP trough and \heii\ forest data.
\Secref{sec:Dis} discusses our results ties together our various measurements to develop a coherent picture of UV background fluctuations, followed by conclusions in 
\autoref{sec:Conc}. Note that, unless otherwise stated, hereafter `\lya' will refer only to the transition due to neutral hydrogen.

\section{Data}\label{sec:Data}
In order to study $2<z<3$ UV background inhomogeneity, we require both observer-frame UV and optical spectra. The UV spectral range gives us access to the \heii\ \lya\ forest, while the optical spectral range gives us the corresponding \hi\ \lya\ forest, and the \ovi\ forest. We must limit ourselves to quasars that fulfil a number of requirements,
\begin{enumerate}
\item the optical spectra must be observed in high signal-to-noise (S/N~$\gtrsim 50$ per pixel) and high spectral resolution (R $\gtrsim 50000$),
\item sufficient quasar continuum emission must remain for \heii\ \lya\ forest absorption analysis such the signal-to-noise is greater than $\sim 0.5$
i.e. the quasar must be a `\heii\ quasar',
\item a significant path in the \heii\ \lya\ forest must be available that does not show a GP trough.
\end{enumerate}
There are only a small number of known \heii\ quasars due to the cumulative impact of \hi\ Lyman limit systems eroding the continuum emission. Of these only \HE\ ($z$\textsubscript{em} = $2.887$) and \HS\ ($z$\textsubscript{em} = $2.748$) fulfil our requirements.
In both spectra, we discard the reddest $\sim 3000$ km~s\textsuperscript{-1} in order to avoid intrinsic quasar effects.

In addition to providing \heii\ forest absorption at $z\lesssim 2.7$, \HE\ includes a GP trough that covers the redshift range, $2.7\lesssim z \lesssim 2.88$ (with some transmission windows). 
Any redshift chunk (of size $1\times 10^{-3}$) was treated as part of a GP trough when more than half the pixels have $\tau_\heii > 5$ (following the method of \citealt{syphers2011}).

\begin{table*}
\caption{Observations Utilized}
\label{tab:data}
\begin{tabular}{ccccccccc}
\hline
QSO     & $z_{em}$  & Instrument      & Exposures (sec)                      & \textlangle S/N/pixel \textrangle     & Resolution       & Notes \\
\hline
\hline
\HE	    & 2.887     & COS G140L/1230  & $1\times 11558$                      & 3                                     & 1500-4000        & GTO 11528 (PI J. Green) \\
{}      & {}        & COS G130M/1222  & $2\times 14814$                      & 8                                     & 16000-21000      & GO 13301(PI J. Shull)   \\
{}      & {}        & \VLT-UVES       & $21600$ $+$ $28800$                  & 49 (\ovi), 121 (\lya)                 & 45000            & \citealt{kim2013}  \\
\hline
\HS 	& $2.748$   & COS G140L/1230  & $1\times 15705$                      & 1                           & 1500-4000        & \citealt{Syphers2013}   \\
{}      & {}        & COS G130M/1222  & $3\times 15694$ $+$ $1\times 12469$  &  1                           & 16000-21000      & GO 13301 (PI J. Shull)  \\
{}      & {}        & \Keck-HIRES     & $2\times 2800$ $+$ $1 \times 3000$   & 61 (\ovi), 83 (\lya)                  & 48000            & C13H (PI W. Sargent)$^1$\\

\hline
\end{tabular}
\begin{description}
\item[] $^1$Retrieved from the KODIAQ database \citep{kodiaq2015,kodiaq2017,Lehner2014}
 \end{description}

\end{table*}
\subsection{\HST\ COS data\label{sec:UVdata}}

The UV spectra were taken by \HST\ COS (\autoref{tab:data}). The G140L/1230 spectrum for \HS\ is the same as analysed in \citet{Syphers2013}. For the G130M and remaining G140L/1230 (for \HE) the extractions followed the same procedures outlined in \citet{Syphers2013}, with the precise details noted below.

While G130M data, as well as newer G140L data exists for \HE\ and \HS, they do not provide the necessary path length. Moreover the higher spectral resolution of G130M has no impact since we filter to cMpc scales. \FUSE\ spectra for \HE\ and \HS\ are available, however, they lacks the required S/N for our analysis. Therefore, we utilise exclusively G140L/1230 in the UV (aside from the exception discussed below).

\begin{figure*}
 \centering
 	\includegraphics[angle=0,width=\linewidth]{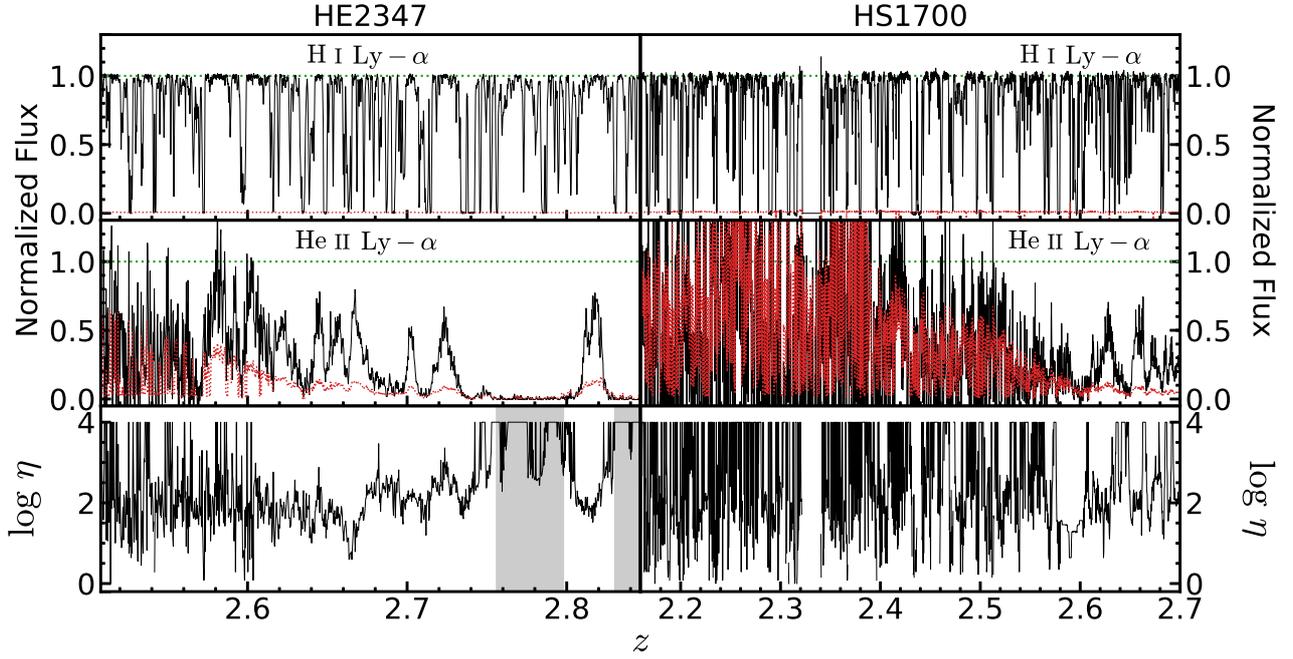}
 \caption{The continuum normalised spectra of \HE\ (left) \& \HS\ (right) for \hi\ \lya\ (Top) and \heii\ \lya\ (Middle), with the flux shown in black, and the $1\sigma$ errors shown as the red dotted lines. The bottom row shows the $\log \eta$ values in black for the full calculated range, with the grey bands denoting the region in the \heii\ GP trough.}
 \label{fig:cont_norm}
 \end{figure*}

In order to continuum normalise the UV spectra, we first identified regions relatively free of absorption and then used an iterative sigma clipping \citep{Syphers2013,Syphers2014} until convergence is met. For \HE, this continuum fit was an extrapolation of the continuum below the \heii\ GP trough with a power-law, $\alpha_\nu$, of -0.4609. For \HS, the extrapolated power-law is given by an $\alpha_\nu$ of -0.1348. The resulting continuum normalised spectra for both are shown in \autoref{fig:cont_norm}.

G140L/1230 has significant wavelength calibration errors in the portion of data of interest here \citep{Syphers2013} due to a lack of usable lines in the onboard Pt/Ne lamp.
\citet{Syphers2013} found that improved calibration was possible by performing a cross-comparison with FUSE data aligning features in the G140L/1230 data with the more reliable wavelength calibration of FUSE. In this work, we use the reliable wavelength solution of G130M data to correct for any remaining calibration errors in both spectra.

\subsection{Optical ground based observations}
The optical spectrum (\autoref{tab:data}) of \HE\ is the same spectrum analysed by \citet{kim2013}, taken with the \VLT\ UVES at the resolution of 6.7~km~s\textsuperscript{-1} and the S/N $\gtrsim 50$ per pixel in the \lya\ forest and \ovi\ region.
We acknowledge that there has been some disagreement on the placement of the continuum \citep{McQuinn2014}, but as our analysis is driven by variation in $\eta$ rather than the overall values, we are broadly resistant to this potential systematic error.

The optical spectrum of the quasar \HS\ (\autoref{tab:data}) comes from the \Keck\ program C13H (PI W. Sargent). The extracted, continuum-normalised, and combined HIRES spectrum was downloaded from the Keck Observatory Database of Ionized Absorbers toward QSOs (KODIAQ) database \citep{kodiaq2015, kodiaq2017, Lehner2014}.

\section{{\texorpdfstring{$\eta$ Forward Modelling}{Eta Forward Modelling}}} \label{sec:etaForwardModeling}
Early work calculating $\eta$ through a direct ratio of opacities was not suited to the resolution and S/N of the available data \citep{McQuinn2014,Syphers2014}.
Others tried to measure $\eta$ as a direct ratio of column densities \citep{Kriss2001, Zheng2004, Fechner2007_461}, but this is limited by degeneracies in line decomposition \citep{Fechner2007_461}.
The modern approach of forward modelling was developed to mitigate these concerns \citep{McQuinn2014,Syphers2014}.

 \begin{figure*}
 \centering
  \includegraphics[angle=0,width=\linewidth]{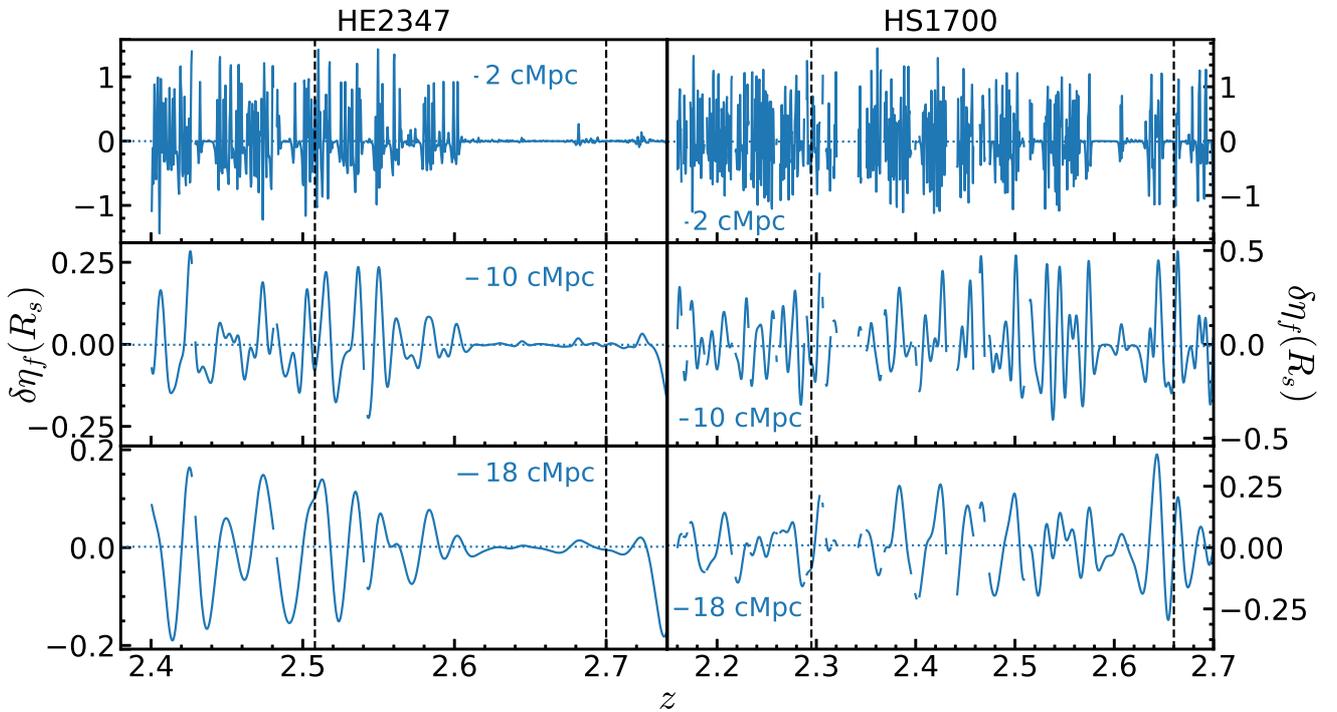}
 \caption{Examples of $\eta$ contrast band-pass filtered on various scales for \HE\ (left) \& \HS\ (right). The panels show (from top to bottom) $R_s=2$~cMpc, $R_s=10$~cMpc, $R_s=18$~cMpc with the relevant scales marked in the caption as the blue line. The redshift paths used for \ovi\ ($2.508<z<2.7$ for \HE\ \& $2.295<z<2.66$ for \HS) are indicated by the vertical  dashed lines.
 }
 \label{fig:smoothmpc2347}
 \end{figure*}

We utilised this method and forward modelled the \heii\ transmission from the \hi\ transmission via a grid of $\eta$ values \citep{Heap2000,Fechner2007_461,McQuinn2014,Syphers2014}. As our spectrum contained no obvious emission lines within the range of interest, we censored any pixels with high normalised flux (flux $>1$) to the normalised value of 1, removing the effect of these noisy pixels from the forward modelling. 

This forward modelling method begins with the censored \hi\ spectrum and a grid of $\eta$ values. This grid runs from $\eta=1$ to 10001, with the intervening grid value $\eta_j$ defined as 
\begin{equation}
\eta_j=\left(\frac{j\sqrt{10001}-1}{j-1}+1\right)^2.
\label{eq:eta_grid}
\end{equation}
Using the definition of $\eta$ (\autoref{eta}), this $\eta$ grid is converted into a set of \heii\ flux grids using complex \hi\ fluxes. These \heii\ fluxes are convolved with the \HST\ COS Line-Spread Function creating a grid of \HST\ resolution model spectra.

The $\eta$ value in redshift bins ($\Delta z=0.0003$) is given by the model spectrum that is the best match to the \HST\ data. If either \hi\ \lya\ or \heii\ \lya\ is undefined for a given pixel, it is flagged to be ignored in later analysis.
For our estimate of $\eta$, we selected the maximum redshift for \HE\ ($z=2.82$) and for \HS\ ($z=2.7$) to avoid proximity effects at $\sim 3000$ km~s\textsuperscript{-1}.
Our minimum redshift is ($z=2.4$ for \HE\ and $z=2.16$ for \HS) is driven by our requirement of S/N~$\gtrsim 50$ per pixel and by avoidance of the Ly-$\beta$ forest.

Our forward modelling gives a median $\eta$ value of $\sim 80$ for \HE\ and $\sim 167$ for \HS\ for the redshift range of our final analysis. While this method addresses some of the issues for the direct determination of $\eta$, it still has some limitations. The largest source of uncertainty is the choice of continuum for the \hi\ spectra (and to a lesser extent \heii), which  affects the estimated $\eta$ \citep{McQuinn2014}. 
The combination of these effects limits the interpretation of the absolute values of $\eta$.
We stress that once more that, in this work, variation in $\eta$ is key and the absolute values are not interpreted.

\subsection{{\texorpdfstring{Filtering $\eta$ on various scales}{Filtering Eta on various scales}}\label{sec:smoothing}}
Our goal in this work is to investigate the strength and physical scale of UV background inhomogeneities. We applied a band-pass filtering to $\eta$ and proceed to study \ovi\ enrichment on this characteristic scale in splits of the filtered field. In order to preform this filtering, we define the contrast in $\eta$ (\autoref{eq:delta_eta})
\begin{equation}
\delta\eta = \frac{\eta-\bar{\eta}}{\bar{\eta}}.
\label{eq:delta_eta}
\end{equation}

We used a Gaussian band-pass filter to extract the forward modelled $\delta\eta$ on various characteristic scales ($R_s$) in comoving megaparsecs (cMpc) using Planck 2015 cosmology evaluated at $z=3$ \citep{Planck2015}. 

Our Gaussian band-pass filter is created using a Gaussian function for the low-pass filter, 
\begin{equation}
f_{G1}(x)=e^{-\frac{1}{2}\left(\frac{2x\sqrt{2\ln{2}}}{R_l}\right)^2},
\label{eq:window}
\end{equation}
(where $R_l$ is the full width half maximum low-pass scale) and a Gaussian function for the high-pass  filter
\begin{equation}
f_{G2}(x)=e^{-\frac{1}{2}\left(\frac{2x\sqrt{2\ln{2}}}{R_h}\right)^2}
\label{eq:window_high}
\end{equation}
(where $R_h$ is the full width half maximum high-pass scale). In both Gaussians, $x$ is defined as the line-of-sight distance in comoving coordinates. The filtering scales were set by
\begin{equation}
R_h=R_s+1\mbox{ cMpc}, R_l=R_s-1\mbox{ cMpc}.
\label{eq:FWHM}
\end{equation}
The two Gaussians and the forward modelled $\eta$, were then transformed into Fourier space 
$\hat{F}(k)=\mathcal{F}\{f(x)\}$ using the FFT command from the scipy fftpack Python 3 package \citep{scipy:2001}. The low-pass filter is simply the Fourier transform of $f_{G1}$,
\begin{equation}
\hat{F}_{low-pass}(k)=\mathcal{F}\{f_{G1}(x)\}.
\label{eq:window_low_fft}
\end{equation}
The high-pass filter is derived taking one minus the Fourier transform of $f_{G2}$ 
\begin{equation}
\hat{F}_{high-pass}(k)=1-\mathcal{F}\{f_{G2}(x)\}.
\label{eq:window_high_fft}
\end{equation}
These filters were then multiplied with the Fourier space $\delta\eta$ and then inverse transformed back in to real space using the iFFT in the same package (\autoref{eq:filter_data}),
\begin{equation}
\delta\eta_f (R_s)=\mathcal{F}^{-1} \big\{\mathcal{F}\{\delta\eta\}\hat{F}_{low-pass} \hat{F}_{high-pass})\big\}.
\label{eq:filter_data}
\end{equation}

In forward modelling $\eta$, some pixels were undefined and these rare instances were interpolated over.
Following the completion of the inverse Fourier transform, the pixels were then re-flagged as bad pixels.
Additionally, the neighbouring pixels were also flagged in a effort to be conservative.
\autoref{fig:smoothmpc2347} shows some example filtered scales for \HE\ and \HS. The amplitude of $\delta\eta_f$ decreases with $z$ for \HE. This is appears to be a consequence of the decreasing noise in the \heii\ forest. However, it is possible to split the sample around the median $\delta\eta_f$ without biasing the analysis.

On $\gtrsim$ 20~cMpc, absorption structure may become degenerate with quasar continuum variation and thus suppressed in a continuum fit. Therefore, scales larger than 20~cMpc along the line-of-sight are unreliable. 
On the other hand, on small scales ($\lesssim 1$~cMpc) thermal line broadening dominates \citep{Fechner2007_461,McQuinn2014}. 
Therefore, we conservatively limited ourselves to bandpass filtering scales in the range 2~cMpc $\lesssim R_s \lesssim $ 20~cMpc.

This band-pass filtering allows us to direct our exploration of UV background to desired scales, while also suppressing the very large or very small scales which may be dominated by systematic errors or undesirable physical effects.

\section{Pixel Optical Depth}\label{sec:POD}
The Pixel Optical Depth (POD) method is a pixel-by-pixel approach that searches for statistical excesses in metal absorption associated with \lya\ absorption. Using this method, originally developed by \citet{Cowie1998}, one determines the \hi\ optical depth for \lya\ forest pixels and pairs them with estimated apparent metal opacity at the same redshift by interpolating at the appropriate wavelength. 

Here we will briefly summarise the key elements to the approach as applied to the \ovi\ absorption. For a detailed account of the method used see \citet{Aguirre2002}. Also see \citet{Cowie1998}, \citet{Schaye2003} and \citet{Pieri2004} for discussion. 
\ovi\ absorption resides fully within the both the \lya\ forest and the Ly-$\beta$ forest and is also affected by other Lyman series lines. Hence some effort must be made to minimise this contaminating absorption. To this end, the doublet nature of \ovi\ absorption can be used to clean the sample of contaminating forest absorption. Here the optical depth is measured for both of the doublet lines, and then the lower equivalent optical depth is chosen (accounting for the difference in oscillator strength and wavelength) as the least contaminated measure. Furthermore, the pixel optical depth approach enables the search for metal lines even when the \lya\ absorption is saturated and so the maximum \lya\ opacity becomes unconstrained.
One can recover a \lya\ opacity measurement using the least contaminated among unsaturated higher-order Lyman series lines (again correcting for differences in oscillator strength and wavelength). We apply this method using 5 Lyman series lines.

We selected a maximum POD analysis redshift of $z=2.70$ for \HE\ to avoid it's GP trough. This limit is a conservative cut, using the same buffer as used for the intrinsic quasar effects from the potential end of the \GP\ trough at $z\sim 2.74$. 
In the case of \HS\ the maximum POD analysis redshift is $z=2.66$ to avoid quasar proximity effects at $\sim 3000$ km~s\textsuperscript{-1} plus a conservative buffer of 3 times the full width at half maximum of our maximum scales.
The minimum redshifts are limited to exclude the quasar \hi\ Ly-$\beta$ emission, though as will be clear below, in reality we further limit this choice.

Armed with best estimates of opacity for the sample of \lya\ and \ovi\ pixel pairs, they are then binned by \lya\ opacity and the median opacity for both \lya\ and \ovi\ in these bins were calculated. Hence the standard POD plot shows $med (\tau_{OVI})$ in bins of $med(\taulya)$. This is shown as a dashed line in \autoref{fig:HE2347_pix_OVI}. A significant correlation with a positive gradient indicates a measurement of correlated metal absorption \citep{Pieri2004}.

We adapted the method of splitting the POD sample used in \citet{Pieri2006}. In that publication the authors split the sample into near/far regions from galaxies on 600 km~s$^{-1}$ scales using either directly observed as Lyman break galaxies or by the use of strong \ion{C}{IV} absorption as a proxy. Here we split our sample of pixels by high/low $\eta$ (among other comparisons) and go a step further by quantifying the ensemble difference in \ovi\ opacity as outlined in \autoref{sec:dT}. Error bars are estimated by bootstrap resampling the observed spectrum from 5\AA\ chunks. Following \citet{Pieri2006} we preserve information on whether pixel pairs fall into the high or low $\eta$ sample in the process. At least 25 pixels from at least 5 unique chunks must be available for a given \lya\ bin in order for us to report the measurement for that bin.

\subsection{\texorpdfstring{\ovi\ as a tracer of ionizing flux}{OVI as a fracer of ionizing flux}}

Various ions can be used to trace the ionizing flux levels at different redshifts. As shown in \citet[Figure 1]{Agofonova2007}, the ionization thresholds of different ions 
probe different portions of the metagalactic ionizing spectrum. 
Using \heii\ absorption as a proxy for UV hardening we can explore how metals are sensitive to associated shape changes in the UV background. 
Since \ovi\ becomes more observable with strong UV hardening, we concentrated our efforts on its analysis. We further limited our minimum redshifts, from those noted above, to $z=2.508$ for \HE\ and $2.295$ for \HS\ to minimize continuum normalisation uncertainty due to the presence of Lyman limit systems in the spectra and their associated Lyman limit breaks.

 \begin{figure*}
 \centering
 \includegraphics[angle=0,width=.49\linewidth]{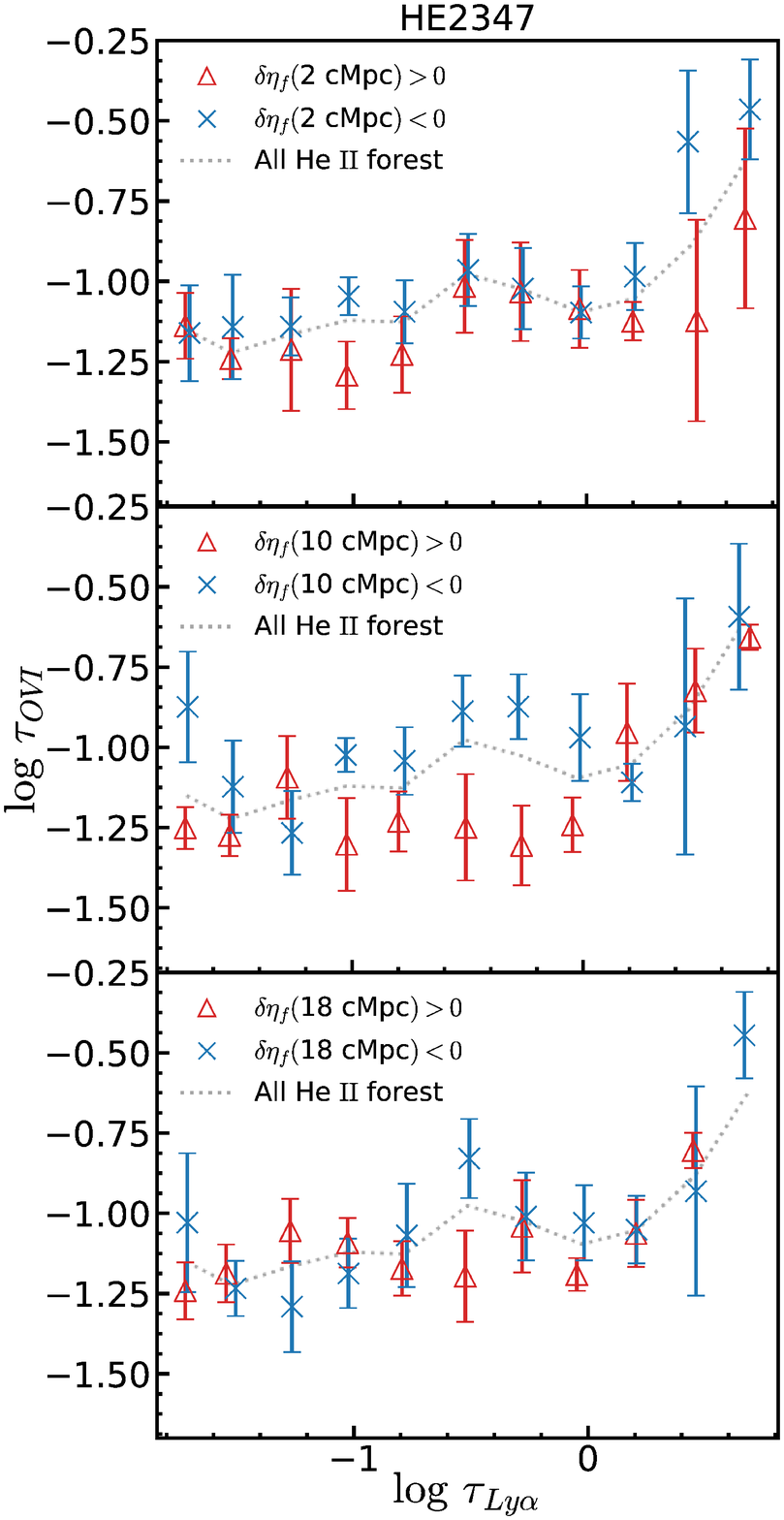}
  \includegraphics[angle=0,width=.49\linewidth]{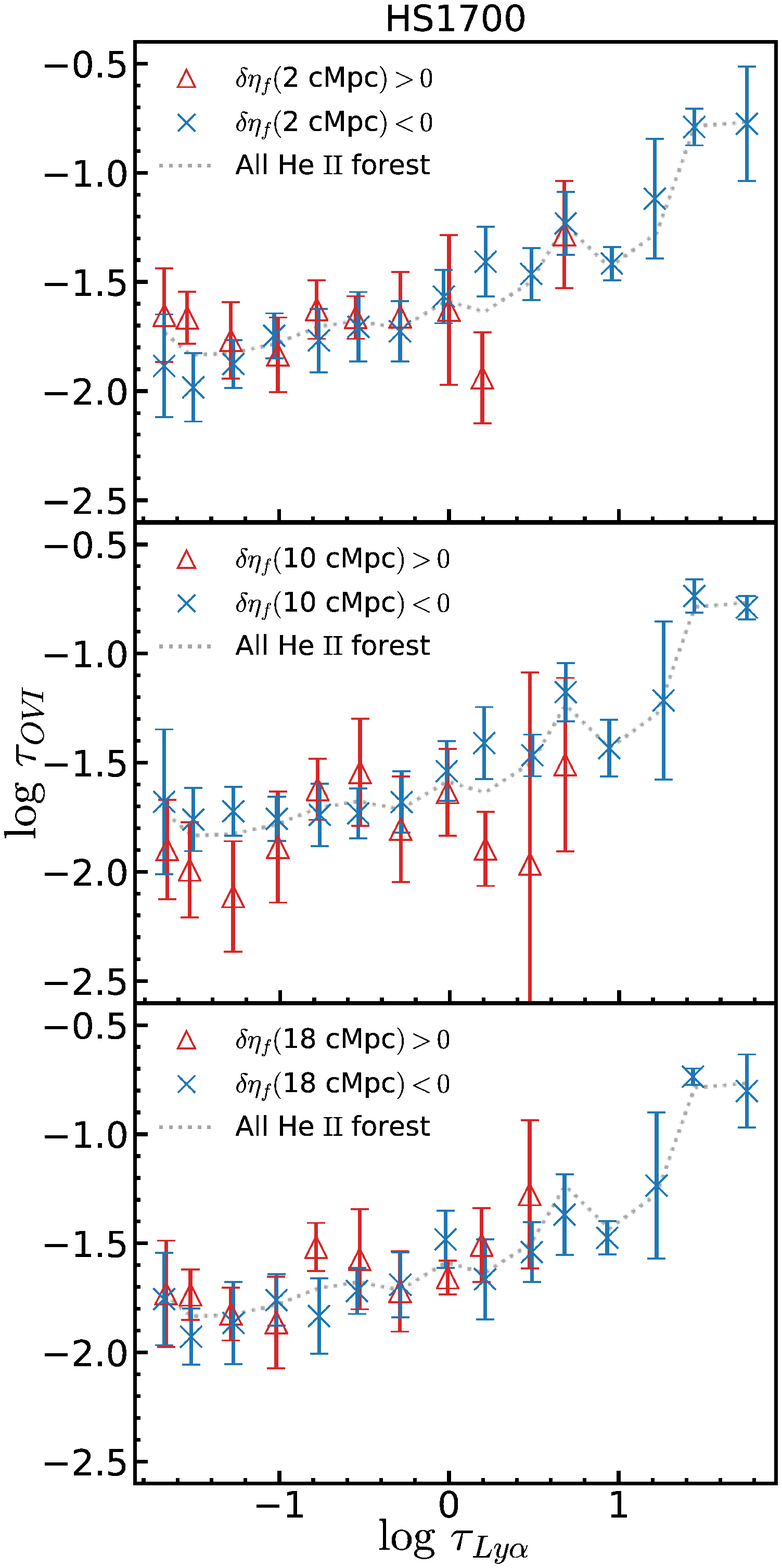}
 \caption{Pixel optical depth measurements for \HE\ (left; $2.508<z<2.7$) \& \HS\ (right; $2.295<z<2.66$) with high-low $\delta\eta$ split on 2~cMpc(top), 10~cMpc(centre), and 18~cMpc(bottom). The high-$\eta$ sample is indicated by the red triangles, and the low-$\eta$ sample is indicated by the blue crosses. 
 For comparison, the combined all \heii\ forest POD results are indicated by the dotted lines. A \ovi\ opacity difference is evident between samples associated with high and low $\eta$ filtered on 10~cMpc scales for \HE\ only.
 } 
 \label{fig:HE2347_pix_OVI}
 \end{figure*}

\autoref{fig:HE2347_pix_OVI} shows the pixel optical depth search for \ovi\ in sub-samples split by high/low $\eta$ filtered on 3 scales (2~cMpc, 10~cMpc and 18~cMpc) for both quasars. For comparison the results of the combined all \heii\ forest pixels POD analysis are also shown (dotted line). The red triangles show the results for the 50\% highest $\eta$ pixels and the blue crosses show the POD analysis for the 50\% lowest $\eta$ pixels. As can be seen in the middle panel for \HE, \ovi\ is sensitive on $\sim$ 10~cMpc scales to this high/low $\eta$ split. This 
demonstrates that large-scale UV background fluctuations as identified by a filtered $\eta$ field and can be probed and confirmed through the measurement of \ovi\ absorption. This is evident on approximately 10~cMpc scales as observed in \HE, but is not present on the smallest or largest scales that we can probe, nor is it seen in the analysis of \HS. We will return to explore this latter point in \autoref{sec:HE_vs_HS}. In the following section we will quantify potential differences in \ovi\ optical depth, assess significance, and explore dependence on physical scale.

\section{Measuring the Differential Oxygen Opacity}\label{sec:dT}

 	\begin{figure*}
	\centering
	\includegraphics[angle=0,width=\linewidth]{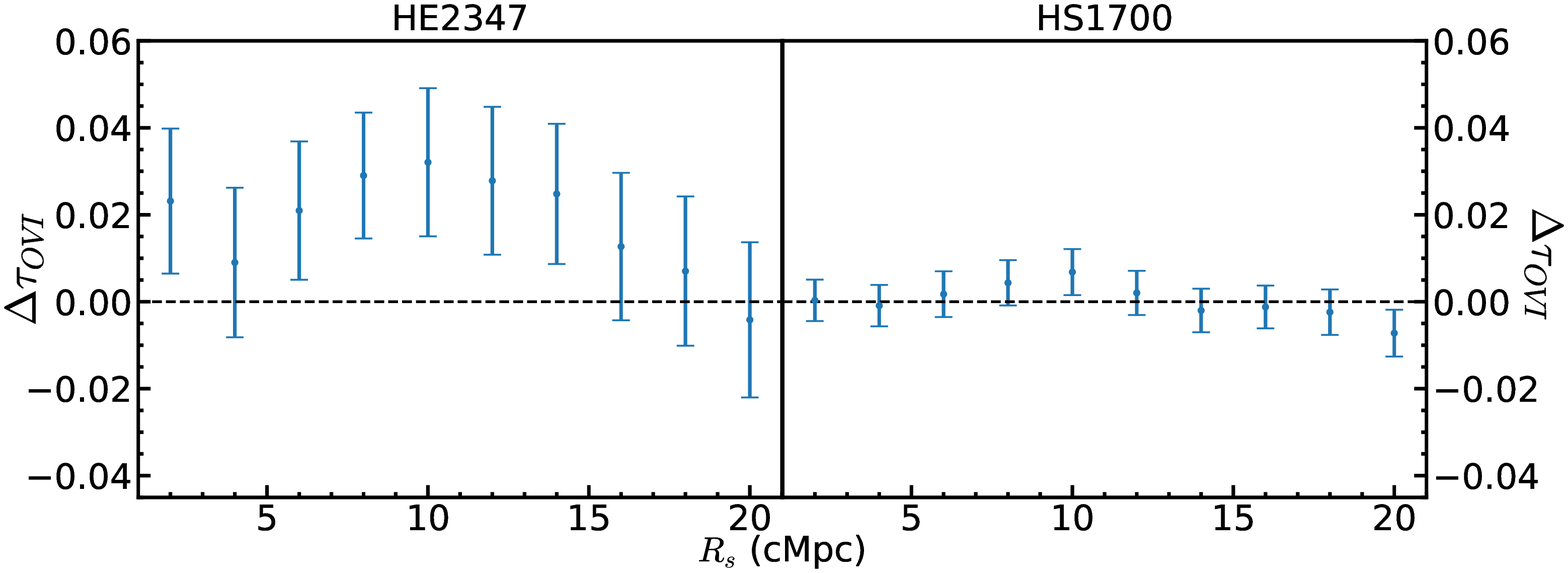}
	\caption{$\Delta\tau_\ovi$ vs filtered scale $R_s$ for \HE\ (left; $2.508<z<2.7$) \& \HS\ (right; $2.295<z<2.66$), where $\Delta\tau_\ovi$ is calculated using a 1-bin POD analysis, matching the standard POD bins with statistics in for both the high- and low-$\eta$ samples. This shows a significant deviation for scales on the order of 10~cMpc for \HE, and no significant deviation for \HS.}
	\label{fig:dTsmooth}
	\end{figure*}

To explore the difference in oxygen absorption between samples (including low and high $\eta$ splits), we now define an extension to the POD approach for this specific purpose. We may calculate an \ovi\ opacity difference
\begin{equation}\label{eq:ovidiff}
\Delta\tau_{OVI} \equiv med(\tau_{OVI, S1}) - med(\tau_{OVI, S2}),
\end{equation}
where subscript $S1$ and subscript $S2$ refer to sample 1 and sample 2 respectively. Where necessary, sample 1 is the sample expected to reflect harder UV conditions and/or lower $\eta$.
In the standard POD approach, the median metal opacity is calculated for a specified bin in $\taulya$. Rather than taking several bins in $\taulya$ as in the standard POD approach (shown in \autoref{fig:HE2347_pix_OVI}),  we calculated $\Delta\tau_{OVI}$ for a single, wide bin reflecting the full range for which $\taulya$ is measurable. 
For each filtered scale, this is defined as the range of standard POD $\taulya$ bins that provide sufficient statistics to support a measure of $\tau_{OVI}$ in {\it both} samples. We utilise this broad $\taulya$ bin in order to simplify the estimation of uncertainty and to avoid unwanted weighting in the opacity difference measurement. The single bin approach is made possible by the fact that, in all the sample splits which follow, the difference in median $\taulya$  between the split samples is insignificant.

\subsection{{\texorpdfstring{Splitting into high/low $\eta$ samples}{Splitting into high/low eta sample}}}\label{sec:highlow}

Here we calculate the \ovi\ absorption difference for pixel samples split by high/low $\eta$ for various filtering scales. Sample 1 in \autoref{eq:ovidiff} refers to the low $\eta$ sample and sample 2 is the high $\eta$ sample. In \autoref{fig:dTsmooth}, this $\Delta\tau_\ovi$ is shown as a function of filtering scale for a 50:50 volumetric split. For \HE, this difference peaks at a scale of 10~cMpc with an \ovi\ excess $\Delta\tau_\ovi=0.032\pm0.017$ with a negligible corresponding difference in median $\taulya$ between these samples (smaller than the standard POD analysis bin size), ruling out the null hypothesis of $\Delta\tau_\ovi=0$ at $2\sigma$. This can be interpreted as the characteristic scale of fluctuations in that line-of-sight. In \HS, there is no significant detection of a characteristic scale. This line-of-sight difference could be interpreted as a large scale fluctuation, larger than can be assessed in a single line-of-sight (see \autoref{sec:HE_vs_HS} and \autoref{sec:Dis}). 
Note that the mean redshift of the high and low $\eta$ samples used in our pixel optical depth analyses are $2.61$ and $2.60$ for \HE, and $2.49$ and $2.50$ for \HS. Therefore, any potential redshift evolution does not affect our results. 

Thus far we have implicitly assumed, through our 50/50 split, that the hard UV background regions occupy an equal volume to soft UV background regions, but this need not be the case. We may relax this assumption and vary the  relative proportions in the two samples. However, the value of such an exploration is limited by the fact that (as pointed out in \autoref{sec:smoothing}) the amplitude of $\delta\eta_f$ is sensitive to the noise level of the \heii\ forest. The special case of a 50/50 split is not vulnerable to this noise dependence (since a sign change in $\delta\eta_f$ will not occur as a result of varying amplitude), but other splits may result in systematic errors. With these caveats in mind we have explored such differences in high/low $\eta$ sample sizes and find no significant difference from the 50/50 split results for either quasar. Hereafter, `high/low $\eta$' will refer only to a 50/50 high/low $\eta$ split.

\subsection{\texorpdfstring{Null tests of high/low $\eta$ splitting}{Null tests of high/low eta splitting}}\label{sec:Null}
In order to confirm that the signal we are seeing it genuine and not just an artefact, we conducted a pixel optical depth measurement at an offset of $\pm$1.5~\AA\ from the \ovi\ doublet (i.e. $\sim 1030.5$ \& $\sim 1033.5$~\AA). We initially chose this offset by selecting spectral regions thought to be bare of correlations identified in \citet{Pieri_etal2014}. We then performed the standard, all pixel, POD analysis to confirm that the POD did indeed return a null result for a single (all pixels) sample.

\begin{figure*}
\centering
\includegraphics[angle=0,width=\linewidth]{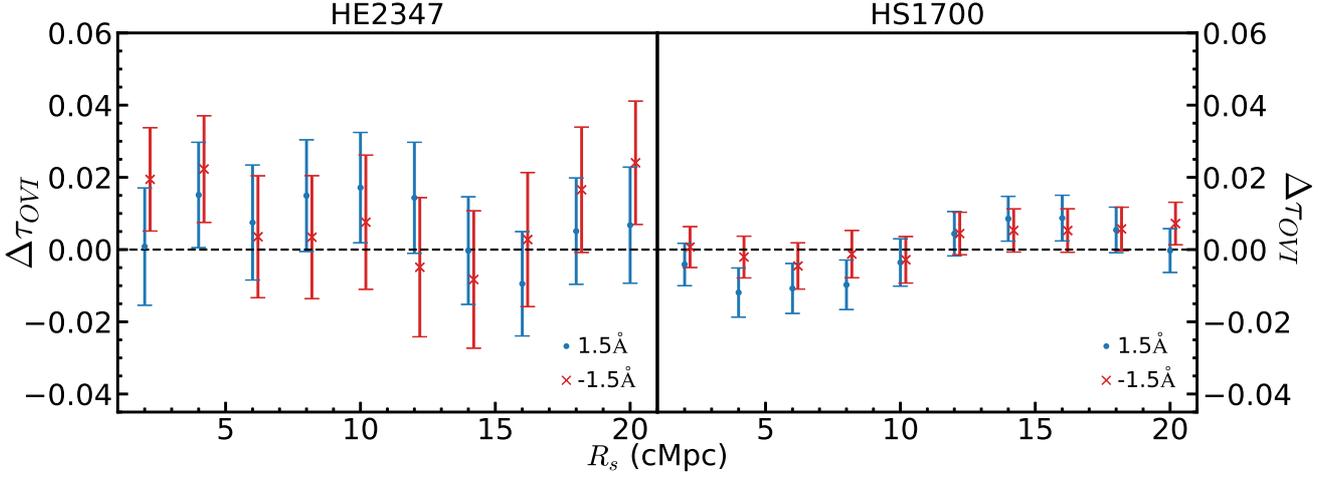}
 \caption{Null test sample of $\Delta\tau_\ovi$ vs filtered Scale ($R_s$ in cMpc) for \HE\ (left; $2.508<z<2.7$) \& \HS\ (right; $2.295<z<2.66$). The analysis was performed by modifying the restframe wavelength of the \ovi\ with offsets of ${+1.5}$\AA\ and ${-1.5}$\AA.}
 \label{fig:Null_dsmoothing}
\end{figure*}

\autoref{fig:Null_dsmoothing} shows the $\Delta\tau$ observable for the null offset for varying filtering scales. Our null result are, as expected, consistent with no signal. Therefore, we find that the signal that we see for \ovi\ is real and not due to a some artefact of the analysis procedure that impacts on local portions of the \ovi\ forest.

\subsection{Comparison of \HE\ \& \HS}\label{sec:HE_vs_HS}
\begin{figure}
 \centering
 \includegraphics[angle=0,width=\linewidth]{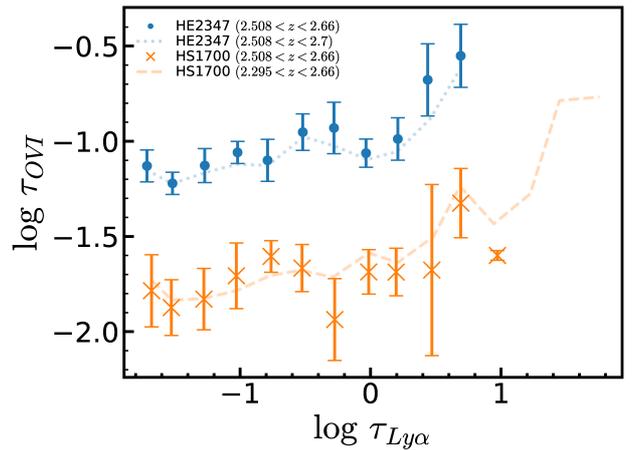}
 \caption{Comparison of all-pixel POD measurements of \ovi\ for \HE\ and \HS\  showing that the difference in signal between the two lines-of-sight large and not due to a redshift dependence. The full path are as the faint lines (\HE\ as the blue dotted and \HS\ as the orange dashed), and the concordance redshift sub-sample is shown as the orange (\HS) and blue (\HE) points.}
 \label{fig:all_pod}
 \end{figure}

In contrast to \HE, \HS\ does not show a strong \ovi\ opacity difference for high/low $\eta$ splits. Additionally, when the all sample \ovi\ POD samples are compared (\autoref{fig:all_pod}), \HS\ contains less \ovi\ than \HE. This holds true even if we shorten the POD paths (285~cMpc for \HE\ and 538~cMpc for \HS) to match the redshift range (a path of 224~cMpc). Following the procedure outlined above for generating a single comparison $\taulya$ bin, we measure $\Delta\tau_\ovi=0.070\pm0.012$ for the matched $2.508<z<2.664$ redshift path. The corresponding difference in median $\taulya$ between these samples is negligible (smaller than the standard POD analysis bin size). This difference between lines-of-sight corresponds to a $6\sigma$ tension: considerably larger than what is seen between the high and low $\eta$ samples shown above. 

As the difference remains even with matching redshift, we cannot attribute the difference seen in $\Delta\tau_\ovi$ to a redshift difference. Rather it appears to be a difference in the UV background, with \HS\ experiencing softer UV radiation than \HE. 

In addition to the differences seen in \ovi\ opacity, there is also a difference seen in in the median $\eta$ values between the samples. 
As noted earlier, the difference exists in the full analysis redshift sample.
This difference becomes slightly smaller with the matched redshift samples, with a median $\eta$ for \HS\ of $\sim 127$ compared to $\sim 74$ seen in \HE. This difference is still significant and reinforces the difference between the two lines-of-sight.

\subsection{\texorpdfstring{Comparison of \ovi\ in the Gunn-Peterson trough and \heii\ forest}{Comparison of OVI in the Gunn-Peterson trough and HeII forest}}\label{sec:GP}
\HE\ shows a significant path in the \heii\ GP trough \citep{shull2010,McQuinn2009}. It is interesting to compare our \heii\ forest data with this GP data and in this section we show the comparison for \HE\ only. \subSecref{sec:discGP} widens the comparison to other samples used in this work and discusses evidence of progress towards complete \heii\ reionization. 

\begin{figure}
\centering
\includegraphics[angle=0,width=\linewidth]{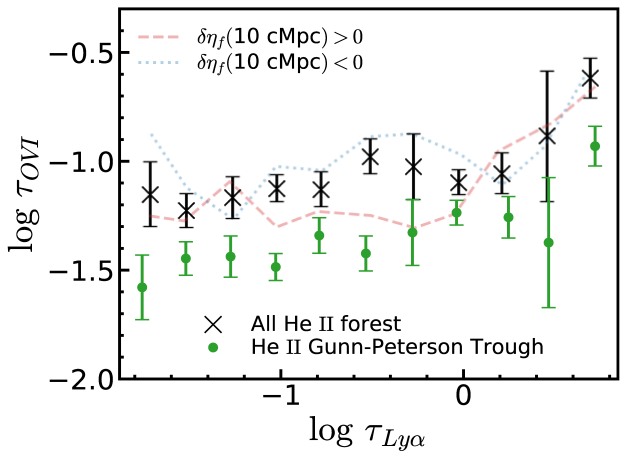}\hfil
 \caption{All POD measurements in the \heii\ GP trough ({\it green points}; $2.7\lesssim z\lesssim 2.88$, as defined in \autoref{sec:Data}) as compared to the all \heii\ forest pixel sample of \HE\ ({\it black crosses}; $2.508<z<2.7$) and the high ({\it red line}) and low ({\it blue line}) $\eta$ ($2.508<z<2.7$) POD measurements previously shown.}
\label{fig:GP}
\end{figure}

\autoref{fig:GP} shows the POD search for \ovi\ absorption for the GP trough data (as defined in \autoref{sec:Data}) and the entire analysis sample of \heii\ forest data in the same spectrum. Also shown for comparison are our main sub-samples from \autoref{sec:highlow}: 10~cMpc filtered $\eta$ split 50/50. There is a clear difference in \ovi\ opacity between the GP pixel sample and the entire \HE\ \heii\ forest sample. Once more measuring this difference in \ovi\ using our standard approach, we derive a measurement of $\Delta\tau_\ovi=0.038\pm0.011$. This is a $3\sigma$ significance measurement of opacity difference. The corresponding difference in median $\taulya$ between these samples is negligible (smaller than the standard POD analysis bin size). The GP sample is marginally consistent with our high-$\eta$ sample, and inconsistent with the low-$\eta$ sample. 

A limitation of the comparison here is the fact that there is a small but significant redshift difference between the GP trough and the \heii\ forest ($\bar{z}=2.60$ and $\bar{z}=2.79$ respectively). Both metallicity evolution and the evolution in the mean flux decrement of the \hi\ \lya\ forest (which enters through residual uncorrected contaminating absorption in the \ovi\ absorption band) could affect our results. However, both these effects are negligible as argued in \autoref{sec:discGP}.

\section{Discussion}\label{sec:Dis}

\subsection{The scale of UV background fluctuations}
In this work we have found significant large-scale differences in \ovi\ opacity associated with fluctuations in $\eta$ on $\gtrsim 5$~cMpc scales seen in the spectrum of \HE. The \ovi\ opacity differences peak at $\Delta\tau_{OVI}=0.032\pm0.017$ on 10~cMpc scales and are significant at the $2\sigma$ level with a negligible corresponding difference in median $\taulya$. 
We find no such signal in the spectrum of \HS\ despite the fact that the measurement precision is higher. However, there is an overall lack of \ovi\ absorption seen in \HS, which is inconsistent with the level of absorption seen in \HE\ at the $6\sigma$ level. The difference between for 224~cMpc paths is $\Delta\tau_{OVI}=0.070 \pm 0.012$ (2.2 times larger than the large scale difference measured within \HE). This combined with the overall higher value of $\eta$ seen in \HS\ indicates that strong $\gtrsim 200$~cMpc UV background fluctuations exist at $z\approx 2.6$. 

We limit ourselves to scales larger than 2~cMpc in order to avoid small-scale systematic effects associated with thermal line broadening \citep{Fechner2007_461,McQuinn2014}.
Our measurements support the assertion that $\eta$ probes real large-scale inhomogeneities in the UV background on several cMpc scales along the line-of-sight (e.g. \citealt{shull2004, Fechner2007_461}).
This is in direct contradiction to the case put forward by \citet{McQuinn2014} based on \HE. In that work, they find weak $\eta$ fluctuations consistent with spurious artefacts and find that their estimated $\eta$ is fully consistent with a homogeneous UV background. We find the most likely explanation of this apparent conflict to be a dilution of their $\eta$ signal due to an over-estimate of the optical continuum in that work (a non-standard solution with essentially no pixels of the forest consistent with zero opacity).

In light of these significant doubts in the community about the degree to which $\eta$ can be treated as a proxy for UV background fluctuations, it has been fundamental to our analysis that we can use \ovi\ absorption to assess it's physical significance. Moreover, we treat scale as a free parameter through band-pass filtering of $\eta$. In this context, our combined analysis of \heii, \hi, and \ovi\ constitutes the first systematic measurement of UV background inhomogenities as a function of scale in the post-helium-reionization universe. We are not able to probe all the potential scales of UV background inhomogeneity, but assess particular windows in scale. We are able to study 2~cMpc $\le R_s \le$ 20~cMpc and $\gtrsim 200$~cMpc scales. We do not attempt to measure $\eta$ on line-of-sight scales larger than 20~cMpc due to concerns that the measured $\eta$ variation will be sensitive to \hi\ continuum normalisation (analogous to how large-scale structures are not measured along lines-of-sight on these larger scales e.g. \citealt{Chabanier2018}). Our results for the fiducial 50/50 case seem, on face value, to indicate that inhomogeneities in the UV background seen at 10~cMpc and are not present on scales of 20~cMpc. However, it is possible that line-of-sight suppression of structure by continuum fitting plays a role erasing $\eta$ fluctuations on scales as small as $\sim$20~cMpc. Hence, we do not consider the absence of fluctuations on 20~cMpc scales to be a reliable result.

Different lines-of-sight can be meaningfully compared since they are not susceptible to common continuum normalisation systematic errors, again analogous to large-scale structure studies (e.g. \citealt{Busca2013}). Also large-scale filtering of $\eta$ is not required. This allows the assessment of line-of-sight scale effects. Conservatively comparing a matched common redshift range between the two quasars equivalent to 224~cMpc, we find a large difference indicating strong inhomogeneities on these scales. This is effectively an integrated quantity since we cannot rule out that our measurement arises due to inhomogeneities on larger scales. Indeed, the entire path of \HS\ appears to be homogeneous within itself and is 538~cMpc in length. 

Our results are consistent with a universe with $\ge 200$~cMpc regions characterised by their soft  or hard  UV background conditions. In this picture, the soft UV regions (\HS) have little or no internal inhomogeneity on 2~cMpc $\le R_s \le$ 20~cMpc scales. The hard UV regions  on the other hand (\HE) do show such inhomogeneity, peaked at 10~cMpc scales. This picture is consistent with all our results and, in particular, brings together our differing results from different quasar spectra without conflict.

Our results are not sensitive to rare strong \ovi\ systems. If the measured signal were a consequence of such outliers, our bootstrap analysis would show errors consistent with a null result. This is unsurprising since the POD analysis is itself outlier resistant through use of the median opacity, and it's pixel-by-pixel approach is a volume average. The standard POD approach may be dominated by such systems for saturated \lya, but our use of single, wide \lya\ opacity bins ensures that our measurements are driven by the majority weaker absorption.

While no other study has systematically measured these scales, various studies provide a useful comparison. \citet{Worseck2006} and \citet{Worseck2007} find that $\eta$ is sensitive to the presence of quasars at transverse separations of a few cMpc, which is consistent with our findings. \HE\ is one of the quasars used in these studies with known proximate quasars, but no quasars have been detected proximate to \HS. This is consistent with our above findings that \HE\ shows systematically more hard UV photons and shows inhomogeneity, while \HS\ appears to be homogeneously soft. Also, \citet{Fechner2007_461} study large-scale $\eta$ fluctuations in both \HE\ and \HS\ along with a small sample of strong \civ\ and \ovi\ systems and find hints of a sensitivity to quasar proximity.

Our measurement of large \ovi\ opacity difference between the two lines-of-sight studied is motivated by our measurements of \heii\ opacity. These results raise the question of whether line-of-sight to line-of-sight variance in \ovi\ opacity have been previously seen in the literature even in the absence of \heii\ \lya\ forest information. While many spectra are available, no one has yet performed a directed analysis on the large-scale coherence length of metal absorption using the POD method. However, the raw POD results shown in figure 7 of \cite{Aracil2004} do appear to show significant line-of-sight to line-of-sight variance at fixed redshift.

\subsection{\texorpdfstring{Probing the end of helium reionization with \ovi\ absorption}{Probing the end of helium reionization with OVI absorption}}\label{sec:discGP}

\HE\ exhibits high redshift \heii\ GP path (while \HS\ does not) and we take the opportunity to compare the \ovi\ absorption in this path with other samples derived from these two spectra.
In \autoref{sec:GP} we presented a comparison to \heii\ forest data in \HE\ and found that  the \heii\ GP trough contains comparable \ovi\ levels to that seen in the soft UV background (high $\eta$) sample. Here we place this GP trough data in a broader context among our measurements.

\begin{figure}
\centering
\includegraphics[angle=0,width=\linewidth]{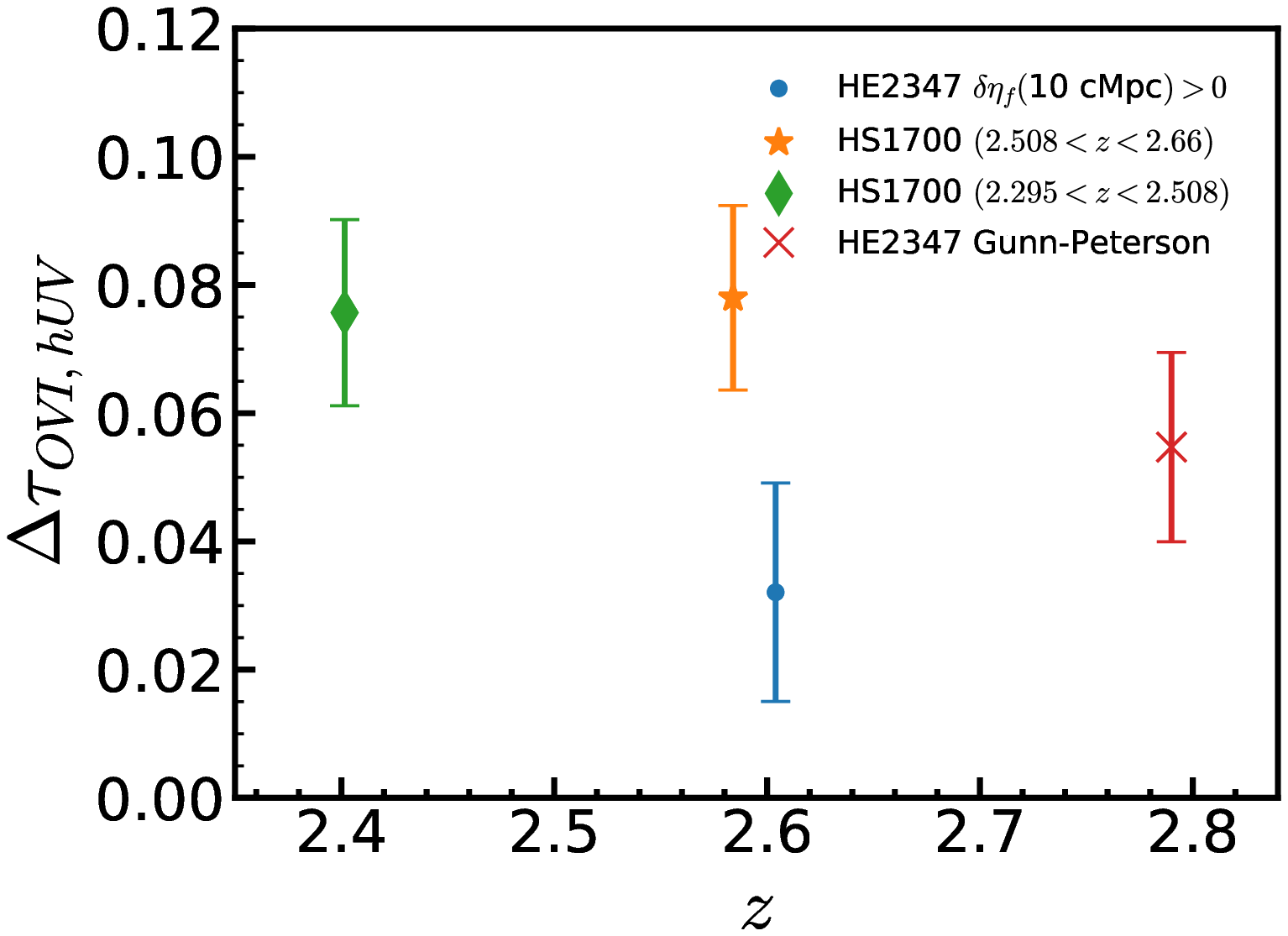}\hfil
 \caption{$\Delta\tau_\ovi$ compared to an `entirely complete reionization' reference sample (see accompanying text). The hard UV background baseline is the 10~cMpc filtered low $\eta$ \HE\ sample and this is compared with the 10~cMpc filtered high $\eta$ sample ({\it blue point}; $2.508<z<2.7$), a high redshift \HS\ sample ({\it orange star}), a low redshift \HS\ sample ({\it green diamond}) and a sample drawn from the GP trough of \HE\ ({\it red cross}; $2.7\lesssim z \lesssim 2.88$, as defined in \autoref{sec:Data}). }
\label{fig:UV_evol}
\end{figure}

\autoref{fig:UV_evol} combines all our \ovi\ difference measurements and compares them to the scale of the difference associated with the \heii\ \GP. We apply once more our standard method for calculating the \ovi\ opacity difference as set out in \autoref{sec:dT}. For this calculation we use a common comparison sample. We select this sample to be the one that shows the hardest UV background conditions and so best reflects {\it an entirely complete \heii\ reionization process}: the low $\eta$ 50\% sample from \HE\ derived from filtering $\eta$ on 10~cMpc scales. We designate this sample `hard UV background sample' and all measurements from this baseline are quoted as
\begin{equation}
\Delta\tau_{\ovi,hUV} \equiv med(\tau_{\ovi, HE,\eta_{low}}) - med(\tau_{\ovi, other}).
\end{equation}
calculated in manner described in \autoref{sec:dT}, where $\tau_{\ovi, HE, \eta_{low}}$ is the baseline sample and $\tau_{\ovi, other}$ is the other sample as indicated in the caption.
It should be noted that, while we characterise this sample as `entirely completed reionization', it is not our lowest redshift sample. Nor do we make any attempts to match redshifts (as we did in \autoref{sec:HE_vs_HS}). Each sample simply uses this as a reference point for reionization progress. For \HS\ we use the narrow redshift comparison sample of \autoref{sec:HE_vs_HS} ($2.508<z<2.66$) and supplement this with the available lower redshift data ($2.295<z<2.508$). The choice to split the \HS\ data by redshift and not by filtered $\eta$ is motivated by the result that splitting \HS\ pixels  yields no significant \ovi\ difference.

It is apparent in \autoref{fig:UV_evol} that no trend towards harder UV background conditions or complete \heii\ reionization is evident in \ovi\ over this narrow, but pivotal redshift range. The most striking result is that the $\bar{z}\approx 2.79$ \heii\ GP trough sample shows no sign of being more deviant from complete reionization than any other sample shown. Extended higher $\eta$ paths such as those seen in a GP trough are expected to reflect softer UV background conditions.
Given the magnitude of our measured \ovi\ sensitivity to $\eta$ outside the GP trough, one would expect a clear signal in \ovi\ if this GP trough were associated with sharp UV background evolution associated with \heii\ reionization. Instead the GP trough \ovi\ absorption from \HE\ is statistically consistent with both the 10~cMpc filtered high $\eta$ data from the same quasar and with \HS\ data despite having a much higher measured $\eta$ than both (see Fig~\ref{fig:cont_norm}).
Overall these results suggest that this \heii\ GP trough does not trace a helium ionization phase change (see following theoretical discussion). 
 
A couple of minor caveats must be noted for completeness at this point. $\Delta\tau_{\ovi,hUV}$ for the red cross and the green diamond in \autoref{fig:UV_evol} may include a contribution from evolution in metallicity or in contaminating forest absorption \citep{Pieri2004}. \cite{Aguirre2008} shows that the redshift evolution probed by the POD approach is weak and smaller than the variance for the small redshift difference here. Contaminating \lya\ absorption can modify the measured apparent \ovi\ opacity in the POD approach, but this also evolves more weakly 
\citep{Kim1997,Kim2007,Faucher-Giguere2008}, 
than the variance between samples. This is further compounded by the fact that the POD approach leaves only a weak residual contribution from forest contamination since efforts are made to reduce such contamination (as described in \autoref{sec:POD}). Overall we find that both these potential systematic affects for the red cross and the green diamond in \autoref{fig:UV_evol} are negligible.

\subsection{Theoretical context}

Our findings are broadly consistent with theoretical predictions of post-\heii-reionization UV background inhomogeneities on $\gtrsim$~10~cMpc scales (e.g. \citealt{Furlanetto2009}). Through calculations of the ionizing background and its connection to helium reionization, \citet{Faucher-Giguere2009} found that the mean free path of \heii\ ionizing photons should be in range of $\sim 15-600$~cMpc. \citet{Davies2017} combined this with their analytic calculation of $\eta$ to perform 3D semi-analytic models of the helium ionizing background. In doing so they put forward a scenario where 200-300~cMpc UV background inhomogeneities and realistic $z<3.2$ GP troughs occur without appealing to \heii\ reionization. Our results favour this scenario. 

These findings have a role in a broader context with implications for the source populations of the UV background and their evolution reaching back to hydrogen reionization. The recent Planck measurements of the opacity of free elections \citep{Planck2016} indicate a relatively late phase of hydrogen reionization, raising the possibility that quasars might be responsible for some or all of this process \citep{Madau2015,Mitra2018, Garaldi2019}. Indeed the measured inhomogeneity in \hi\ absorption at end of hydrogen reionization may require that rare sources such as quasars play an important role \citep{Chardin2017, DAloisio2018, Becker2018}. Furthermore, there has been discussion in the literature of a hitherto unidentified faint and high redshift quasar population \citep{Giallongo2015, Matsuoka2018}. The prospect of an early and extensive contribution from quasars driven by these factors has the potential to conflict with \heii\ GP measurements \citep{Garaldi2019}. While tuned models can bring into agreement \heii\ reionization and a significant early post-\hi-reionization role for quasars \citep{Mitra2018}, a weakening of the perceived connection between \heii\ GP troughs and \heii\ reionization would also ease the apparent tension. Our results of \heii\ GP trough data (studied in the broader context of \ovi\ and its association with UV background fluctuations) favour this weakening of the connection to \heii\ reionization.

\section{Conclusion}\label{sec:Conc}

We have combined \heii\ and \hi\ \lya\ forests, though the ratio $\eta$, along with pixel optical depth measurements of \ovi\ to search for large-scale inhomogeneities in the extragalactic UV background in the sightlines towards \HE\ and \HS. Additionally, we have compared the \heii\ forest lines-of-sight, and the Gunn-Peterson tough in \HE, to search for larger scales inhomogeneities and examine the end of \heii\ reionization. We can summarise our results 
as follows.
\begin{enumerate}
    \item \ovi\ is sensitive to $\eta$ variation filtered on large scales ($\gtrsim 5$~cMpc).
    \item Variation in $\eta$ filtered on large-scales can probe the UV background and is not overwhelmed by observing noise and systematic errors.
    \item  On $\sim 10$~cMpc scales (probed by $\eta$ variation) the UV background varies such that associated median \ovi\ opacity differs by $\Delta\tau_{OVI}=0.032\pm0.017$.
    \item On $\gtrsim 200$~cMpc scales (comparing lines-of-sight) the UV background varies such that associated median \ovi\ opacity differs by $\Delta\tau_{OVI}=0.070 \pm 0.012$
    \item \ovi\ absorption associated with the \heii\ GP trough in \HE\ differs from the \heii\ forest  of the same quasar by $\Delta\tau_\ovi=0.038\pm0.011$ and is broadly consistent with the high $\eta$, soft UV background, path for the same quasar.
    \item \ovi\ absorption associated with the GP trough shows no evolution when compared with samples thought to represent (more) complete \heii\ reionization at lower redshifts, suggesting that this \heii\ GP trough does not trace a sharp \heii\ reionization phase change. 
\end{enumerate}

Overall we find that our analysis favours large-scale ($\sim 10$~cMpc and $\gtrsim 200$~cMpc) UV background fluctuations consistent with recent theoretical projections. Our results are suggestive of extended soft/hard UV background regions, with hard regions showing significant internal inhomogeneities on $\sim 10$~cMpc scales, but soft UV regions lacking such structure. We find no evidence of a contribution from \heii\ reionization in our $z<3$ sample (despite the presence of a GP trough). There are no detailed projections of the magnitude of expected \ovi\ opacity difference with which to compare and so we provide our measured opacity differences to allow such a comparison in future simulation analyses.

Given the limited number \heii\ quasars sufficiently bright for POD analysis, the prospects to significantly extend this analysis to larger datasets are not promising. However, we have established that \ovi\ opacity measured over large coherent paths in the $z\sim3$ \lya\ forest can usefully probe the UV background. Therefore the potential exists to use large-scale \ovi\ inhomogeneity as a proxy for UV background fluctuations, in the absence of \heii\ information. The prospects for direct detection of 
large volumes of the IGM
proximate to transverse quasars is encouraging given the increasing number density of quasars in recent massive spectroscopic surveys Baryon Oscillation Spectroscopic Survey (BOSS; \citealt{Dawson2013}) and extended-BOSS (eBOSS; \citealt{Dawson2016}), and future surveys WEAVE-QSO (\citealt{Pieri2016}) and Dark Energy Spectroscopic Instrument (DESI; \citealt{DESI2016}). Metal absorption and its sensitivity to UV background fluctuations due to detected quasar placement may be probed using stacking methods \citep{Pieri2014} and modified pixel optical depth methods \citep{Pieri2010} in these large and growing surveys.

\section*{Acknowledgements}
We thank Andrea Ferrara, Charles Danforth and Mike Shull for stimulating discussions. 
SM and MP were supported by the A*MIDEX project (ANR-11-IDEX-0001-02) funded by the ``Investissements d'Avenir'' French Government program, managed by the French National Research Agency (ANR), and by ANR under contract ANR-14-ACHN-0021.

This research has made use of the Keck Observatory Archive (KOA), which is operated by the W. M. Keck Observatory and the NASA Exoplanet Science Institute (NExScI), under contract with the National Aeronautics and Space Administration.\
Some/all the data presented in this work were obtained from the Keck Observatory Database of Ionized Absorbers toward QSOs (KODIAQ), which was funded through NASA ADAP grant NNX10AE84G




\bibliographystyle{mnras}
\bibliography{Eta_morrison.bib} 

\begin{thebibliography}{}
\makeatletter
\relax
\def\mn@urlcharsother{\let\do\@makeother \do\$\do\&\do\#\do\^\do\_\do\%\do\~}
\def\mn@doi{\begingroup\mn@urlcharsother \@ifnextchar [ {\mn@doi@}
  {\mn@doi@[]}}
\def\mn@doi@[#1]#2{\def\@tempa{#1}\ifx\@tempa\@empty \href
  {http://dx.doi.org/#2} {doi:#2}\else \href {http://dx.doi.org/#2} {#1}\fi
  \endgroup}
\def\mn@eprint#1#2{\mn@eprint@#1:#2::\@nil}
\def\mn@eprint@arXiv#1{\href {http://arxiv.org/abs/#1} {{\tt arXiv:#1}}}
\def\mn@eprint@dblp#1{\href {http://dblp.uni-trier.de/rec/bibtex/#1.xml}
  {dblp:#1}}
\def\mn@eprint@#1:#2:#3:#4\@nil{\def\@tempa {#1}\def\@tempb {#2}\def\@tempc
  {#3}\ifx \@tempc \@empty \let \@tempc \@tempb \let \@tempb \@tempa \fi \ifx
  \@tempb \@empty \def\@tempb {arXiv}\fi \@ifundefined
  {mn@eprint@\@tempb}{\@tempb:\@tempc}{\expandafter \expandafter \csname
  mn@eprint@\@tempb\endcsname \expandafter{\@tempc}}}

\bibitem[\protect\citeauthoryear{{Agafonova}, {Levshakov}, {Reimers},
  {Fechner}, {Tytler}, {Simcoe}  \& {Songaila}}{{Agafonova}
  et~al.}{2007}]{Agofonova2007}
{Agafonova} I.~I.,  {Levshakov} S.~A.,  {Reimers} D.,  {Fechner} C.,  {Tytler}
  D.,  {Simcoe} R.~A.,   {Songaila} A.,  2007, \mn@doi [\aap]
  {10.1051/0004-6361:20065721}, \href
  {http://adsabs.harvard.edu/abs/2007A%26A...461..893A} {461, 893}

\bibitem[\protect\citeauthoryear{{Aguirre}, {Schaye}  \& {Theuns}}{{Aguirre}
  et~al.}{2002}]{Aguirre2002}
{Aguirre} A.,  {Schaye} J.,   {Theuns} T.,  2002, \mn@doi [\apj]
  {10.1086/341580}, \href {http://adsabs.harvard.edu/abs/2002ApJ...576....1A}
  {576, 1}

\bibitem[\protect\citeauthoryear{{Aguirre}, {Dow-Hygelund}, {Schaye}  \&
  {Theuns}}{{Aguirre} et~al.}{2008}]{Aguirre2008}
{Aguirre} A.,  {Dow-Hygelund} C.,  {Schaye} J.,   {Theuns} T.,  2008, \mn@doi
  [\apj] {10.1086/592554}, \href
  {http://adsabs.harvard.edu/abs/2008ApJ...689..851A} {689, 851}

\bibitem[\protect\citeauthoryear{{Aracil}, {Petitjean}, {Pichon}  \&
  {Bergeron}}{{Aracil} et~al.}{2004}]{Aracil2004}
{Aracil} B.,  {Petitjean} P.,  {Pichon} C.,   {Bergeron} J.,  2004, \mn@doi
  [\aap] {10.1051/0004-6361:20034346}, \href
  {https://ui.adsabs.harvard.edu/\#abs/2004A&A...419..811A} {419, 811}

\bibitem[\protect\citeauthoryear{{Becker}, {Davies}, {Furlanetto}, {Malkan},
  {Boera}  \& {Douglass}}{{Becker} et~al.}{2018}]{Becker2018}
{Becker} G.~D.,  {Davies} F.~B.,  {Furlanetto} S.~R.,  {Malkan} M.~A.,  {Boera}
  E.,   {Douglass} C.,  2018, \mn@doi [\apj] {10.3847/1538-4357/aacc73}, \href
  {http://adsabs.harvard.edu/abs/2018ApJ...863...92B} {863, 92}

\bibitem[\protect\citeauthoryear{{Bolton} \& {Viel}}{{Bolton} \&
  {Viel}}{2011}]{Bolton2011}
{Bolton} J.~S.,  {Viel} M.,  2011, \mn@doi [\mnras]
  {10.1111/j.1365-2966.2011.18384.x}, \href
  {http://adsabs.harvard.edu/abs/2011MNRAS.414..241B} {414, 241}

\bibitem[\protect\citeauthoryear{{Busca} et~al.,}{{Busca}
  et~al.}{2013}]{Busca2013}
{Busca} N.~G.,  et~al., 2013, \mn@doi [\aap] {10.1051/0004-6361/201220724},
  \href {http://adsabs.harvard.edu/abs/2013A%26A...552A..96B} {552, A96}

\bibitem[\protect\citeauthoryear{{Chabanier} et~al.,}{{Chabanier}
  et~al.}{2018}]{Chabanier2018}
{Chabanier} S.,  et~al., 2018, arXiv e-prints, \href
  {http://adsabs.harvard.edu/abs/2018arXiv181203554C} {}

\bibitem[\protect\citeauthoryear{{Chardin}, {Puchwein}  \&
  {Haehnelt}}{{Chardin} et~al.}{2017}]{Chardin2017}
{Chardin} J.,  {Puchwein} E.,   {Haehnelt} M.~G.,  2017, \mn@doi [\mnras]
  {10.1093/mnras/stw2943}, \href
  {http://adsabs.harvard.edu/abs/2017MNRAS.465.3429C} {465, 3429}

\bibitem[\protect\citeauthoryear{{Cowie} \& {Songaila}}{{Cowie} \&
  {Songaila}}{1998}]{Cowie1998}
{Cowie} L.~L.,  {Songaila} A.,  1998, \mn@doi [\nat] {10.1038/27845}, \href
  {http://adsabs.harvard.edu/abs/1998Natur.394...44C} {394, 44}

\bibitem[\protect\citeauthoryear{{Croft}}{{Croft}}{2004}]{Croft2004}
{Croft} R. A.~C.,  2004, \mn@doi [\apj] {10.1086/421839}, \href
  {https://ui-adsabs-harvard-edu.insu.bib.cnrs.fr/#abs/2004ApJ...610..642C}
  {610, 642}

\bibitem[\protect\citeauthoryear{{Croft}, {Weinberg}, {Katz}  \&
  {Hernquist}}{{Croft} et~al.}{1998}]{Croft1998}
{Croft} R. A.~C.,  {Weinberg} D.~H.,  {Katz} N.,   {Hernquist} L.,  1998,
  \mn@doi [\apj] {10.1086/305289}, \href
  {https://ui-adsabs-harvard-edu.insu.bib.cnrs.fr/#abs/1998ApJ...495...44C}
  {495, 44}

\bibitem[\protect\citeauthoryear{{D'Aloisio}, {McQuinn}, {Davies}  \&
  {Furlanetto}}{{D'Aloisio} et~al.}{2018}]{DAloisio2018}
{D'Aloisio} A.,  {McQuinn} M.,  {Davies} F.~B.,   {Furlanetto} S.~R.,  2018,
  \mn@doi [\mnras] {10.1093/mnras/stx2341}, \href
  {http://adsabs.harvard.edu/abs/2018MNRAS.473..560D} {473, 560}

\bibitem[\protect\citeauthoryear{{DESI Collaboration} et~al.,}{{DESI
  Collaboration} et~al.}{2016}]{DESI2016}
{DESI Collaboration} et~al., 2016, arXiv e-prints, \href
  {http://adsabs.harvard.edu/abs/2016arXiv161100036D} {}

\bibitem[\protect\citeauthoryear{{Davies} \& {Furlanetto}}{{Davies} \&
  {Furlanetto}}{2014}]{Davies2014}
{Davies} F.~B.,  {Furlanetto} S.~R.,  2014, \mn@doi [\mnras]
  {10.1093/mnras/stt1911}, \href
  {https://ui-adsabs-harvard-edu.insu.bib.cnrs.fr/#abs/2014MNRAS.437.1141D}
  {437, 1141}

\bibitem[\protect\citeauthoryear{Davies, Furlanetto  \& Dixon}{Davies
  et~al.}{2017}]{Davies2017}
Davies F.~B.,  Furlanetto S.~R.,   Dixon K.~L.,  2017, \mnras, 465, 2886

\bibitem[\protect\citeauthoryear{{Dawson} et~al.,}{{Dawson}
  et~al.}{2013}]{Dawson2013}
{Dawson} K.~S.,  et~al., 2013, \mn@doi [\aj] {10.1088/0004-6256/145/1/10},
  \href {http://adsabs.harvard.edu/abs/2013AJ....145...10D} {145, 10}

\bibitem[\protect\citeauthoryear{{Dawson} et~al.,}{{Dawson}
  et~al.}{2016}]{Dawson2016}
{Dawson} K.~S.,  et~al., 2016, \mn@doi [\aj] {10.3847/0004-6256/151/2/44},
  \href {http://adsabs.harvard.edu/abs/2016AJ....151...44D} {151, 44}

\bibitem[\protect\citeauthoryear{{Fan} et~al.,}{{Fan} et~al.}{2006}]{Fan2006}
{Fan} X.,  et~al., 2006, \mn@doi [\aj] {10.1086/504836}, \href
  {https://ui-adsabs-harvard-edu.insu.bib.cnrs.fr/\#abs/2006AJ....132..117F}
  {132, 117}

\bibitem[\protect\citeauthoryear{{Faucher-Gigu{\`e}re}, {Prochaska}, {Lidz},
  {Hernquist}  \& {Zaldarriaga}}{{Faucher-Gigu{\`e}re}
  et~al.}{2008}]{Faucher-Giguere2008}
{Faucher-Gigu{\`e}re} C.-A.,  {Prochaska} J.~X.,  {Lidz} A.,  {Hernquist} L.,
  {Zaldarriaga} M.,  2008, \mn@doi [\apj] {10.1086/588648}, \href
  {http://adsabs.harvard.edu/abs/2008ApJ...681..831F} {681, 831}

\bibitem[\protect\citeauthoryear{Faucher-Gigu{\`e}re, Lidz, Zaldarriaga  \&
  Hernquist}{Faucher-Gigu{\`e}re et~al.}{2009}]{Faucher-Giguere2009}
Faucher-Gigu{\`e}re C.-A.,  Lidz A.,  Zaldarriaga M.,   Hernquist L.,  2009,
  \apj, 703, 1416

\bibitem[\protect\citeauthoryear{Fechner \& Reimers}{Fechner \&
  Reimers}{2007}]{Fechner2007_461}
Fechner C.,  Reimers D.,  2007, \aap, 461, 847

\bibitem[\protect\citeauthoryear{{Furlanetto}}{{Furlanetto}}{2009}]{Furlanetto2009}
{Furlanetto} S.~R.,  2009, \mn@doi [\apj] {10.1088/0004-637X/703/1/702}, \href
  {http://adsabs.harvard.edu/abs/2009ApJ...703..702F} {703, 702}

\bibitem[\protect\citeauthoryear{Furlanetto \& Oh}{Furlanetto \&
  Oh}{2008}]{Furlanetto2008}
Furlanetto S.~R.,  Oh S.~P.,  2008, \apj, 681, 1

\bibitem[\protect\citeauthoryear{Garaldi, Compostella  \& Porciani}{Garaldi
  et~al.}{2019}]{Garaldi2019}
Garaldi E.,  Compostella M.,   Porciani C.,  2019, \mnras, 483, 5301

\bibitem[\protect\citeauthoryear{Giallongo et~al.,}{Giallongo
  et~al.}{2015}]{Giallongo2015}
Giallongo E.,  et~al., 2015, \aap, 578, A83

\bibitem[\protect\citeauthoryear{Graziani, Maselli  \& Maio}{Graziani
  et~al.}{2019}]{Graziani2019}
Graziani L.,  Maselli A.,   Maio U.,  2019, \mnras, 482, L112

\bibitem[\protect\citeauthoryear{Heap, Williger, Smette, Hubeny, Sahu, Jenkins,
  Tripp  \& Winkler}{Heap et~al.}{2000}]{Heap2000}
Heap S.~R.,  Williger G.~M.,  Smette A.,  Hubeny I.,  Sahu M.~S.,  Jenkins
  E.~B.,  Tripp T.~M.,   Winkler J.~N.,  2000, \apj, 534, 69

\bibitem[\protect\citeauthoryear{Jones, Oliphant, Peterson  et~al.}{Jones
  et~al.}{2001}]{scipy:2001}
Jones E.,  Oliphant T.,  Peterson P.,   et~al., 2001, {SciPy}: Open source
  scientific tools for {Python}, \url {http://www.scipy.org/}

\bibitem[\protect\citeauthoryear{{Kim}, {Hu}, {Cowie}  \& {Songaila}}{{Kim}
  et~al.}{1997}]{Kim1997}
{Kim} T.-S.,  {Hu} E.~M.,  {Cowie} L.~L.,   {Songaila} A.,  1997, \mn@doi [\aj]
  {10.1086/118446}, \href {http://adsabs.harvard.edu/abs/1997AJ....114....1K}
  {114, 1}

\bibitem[\protect\citeauthoryear{{Kim}, {Bolton}, {Viel}, {Haehnelt}  \&
  {Carswell}}{{Kim} et~al.}{2007}]{Kim2007}
{Kim} T.-S.,  {Bolton} J.~S.,  {Viel} M.,  {Haehnelt} M.~G.,   {Carswell}
  R.~F.,  2007, \mn@doi [\mnras] {10.1111/j.1365-2966.2007.12406.x}, \href
  {http://adsabs.harvard.edu/abs/2007MNRAS.382.1657K} {382, 1657}

\bibitem[\protect\citeauthoryear{{Kim}, {Partl}, {Carswell}  \&
  {M{\"u}ller}}{{Kim} et~al.}{2013}]{kim2013}
{Kim} T.-S.,  {Partl} A.~M.,  {Carswell} R.~F.,   {M{\"u}ller} V.,  2013,
  \mn@doi [\aap] {10.1051/0004-6361/201220042}, \href
  {http://adsabs.harvard.edu/abs/2013A%26A...552A..77K} {552, A77}

\bibitem[\protect\citeauthoryear{Kramida, Ralchenko, Reader  \& {and NIST ASD
  Team}}{Kramida et~al.}{2018}]{NIST_ASD}
Kramida A.,  Ralchenko {\relax{}Yu}.,  Reader J.,   {and NIST ASD Team} 2018,
  {NIST Atomic Spectra Database (ver. 5.5.6), [Online]. Available:
  {\tt{https://physics.nist.gov/asd}} [2018, April 12]. National Institute of
  Standards and Technology, Gaithersburg, MD.}

\bibitem[\protect\citeauthoryear{Kriss et~al.,}{Kriss et~al.}{2001}]{Kriss2001}
Kriss G.~A.,  et~al., 2001, \sci, 293, 1112

\bibitem[\protect\citeauthoryear{{Lehner}, {O'Meara}, {Fox}, {Howk},
  {Prochaska}, {Burns}  \& {Armstrong}}{{Lehner} et~al.}{2014}]{Lehner2014}
{Lehner} N.,  {O'Meara} J.~M.,  {Fox} A.~J.,  {Howk} J.~C.,  {Prochaska} J.~X.,
   {Burns} V.,   {Armstrong} A.~A.,  2014, \mn@doi [\apj]
  {10.1088/0004-637X/788/2/119}, \href
  {http://adsabs.harvard.edu/abs/2014ApJ...788..119L} {788, 119}

\bibitem[\protect\citeauthoryear{{Madau} \& {Haardt}}{{Madau} \&
  {Haardt}}{2015}]{Madau2015}
{Madau} P.,  {Haardt} F.,  2015, \mn@doi [\apjl] {10.1088/2041-8205/813/1/L8},
  \href {http://adsabs.harvard.edu/abs/2015ApJ...813L...8M} {813, L8}

\bibitem[\protect\citeauthoryear{Matsuoka et~al.,}{Matsuoka
  et~al.}{2018}]{Matsuoka2018}
Matsuoka Y.,  et~al., 2018, \apj, 869, 150

\bibitem[\protect\citeauthoryear{{McQuinn}}{{McQuinn}}{2009}]{McQuinn2009}
{McQuinn} M.,  2009, \mn@doi [\apjl] {10.1088/0004-637X/704/2/L89}, \href
  {http://adsabs.harvard.edu/abs/2009ApJ...704L..89M} {704, L89}

\bibitem[\protect\citeauthoryear{{McQuinn} \& {Worseck}}{{McQuinn} \&
  {Worseck}}{2014}]{McQuinn2014}
{McQuinn} M.,  {Worseck} G.,  2014, \mn@doi [\mnras] {10.1093/mnras/stu242},
  \href {http://adsabs.harvard.edu/abs/2014MNRAS.440.2406M} {440, 2406}

\bibitem[\protect\citeauthoryear{Mitra, Choudhury  \& Ferrara}{Mitra
  et~al.}{2018}]{Mitra2018}
Mitra S.,  Choudhury T.~R.,   Ferrara A.,  2018, \mnras, 473, 1416

\bibitem[\protect\citeauthoryear{{O'Meara} et~al.,}{{O'Meara}
  et~al.}{2015}]{kodiaq2015}
{O'Meara} J.~M.,  et~al., 2015, \mn@doi [\aj] {10.1088/0004-6256/150/4/111},
  \href {http://adsabs.harvard.edu/abs/2015AJ....150..111O} {150, 111}

\bibitem[\protect\citeauthoryear{{O'Meara}, {Lehner}, {Howk}, {Prochaska},
  {Fox}, {Peeples}, {Tumlinson}  \& {O'Shea}}{{O'Meara}
  et~al.}{2017}]{kodiaq2017}
{O'Meara} J.~M.,  {Lehner} N.,  {Howk} J.~C.,  {Prochaska} J.~X.,  {Fox} A.~J.,
   {Peeples} M.~S.,  {Tumlinson} J.,   {O'Shea} B.~W.,  2017, \mn@doi [\aj]
  {10.3847/1538-3881/aa82b8}, \href
  {http://adsabs.harvard.edu/abs/2017AJ....154..114O} {154, 114}

\bibitem[\protect\citeauthoryear{Oppenheimer \& Schaye}{Oppenheimer \&
  Schaye}{2013}]{Oppenheimer2013}
Oppenheimer B.~D.,  Schaye J.,  2013, \mnras, 434, 1063

\bibitem[\protect\citeauthoryear{Oppenheimer, Segers, Schaye, Richings  \&
  Crain}{Oppenheimer et~al.}{2017}]{Oppenheimer2017}
Oppenheimer B.~D.,  Segers M.,  Schaye J.,  Richings A.~J.,   Crain R.~A.,
  2017, \mnras, 474, 4740

\bibitem[\protect\citeauthoryear{{Pieri} \& {Haehnelt}}{{Pieri} \&
  {Haehnelt}}{2004}]{Pieri2004}
{Pieri} M.~M.,  {Haehnelt} M.~G.,  2004, \mn@doi [\mnras]
  {10.1111/j.1365-2966.2004.07278.x}, \href
  {http://adsabs.harvard.edu/abs/2004MNRAS.347..985P} {347, 985}

\bibitem[\protect\citeauthoryear{{Pieri}, {Schaye}  \& {Aguirre}}{{Pieri}
  et~al.}{2006}]{Pieri2006}
{Pieri} M.~M.,  {Schaye} J.,   {Aguirre} A.,  2006, \mn@doi [\apj]
  {10.1086/498738}, \href
  {https://ui.adsabs.harvard.edu/#abs/2006ApJ...638...45P} {638, 45}

\bibitem[\protect\citeauthoryear{{Pieri}, {Frank}, {Mathur}, {Weinberg}, {York}
   \& {Oppenheimer}}{{Pieri} et~al.}{2010}]{Pieri2010}
{Pieri} M.~M.,  {Frank} S.,  {Mathur} S.,  {Weinberg} D.~H.,  {York} D.~G.,
  {Oppenheimer} B.~D.,  2010, \mn@doi [\apj] {10.1088/0004-637X/716/2/1084},
  \href {http://adsabs.harvard.edu/abs/2010ApJ...716.1084P} {716, 1084}

\bibitem[\protect\citeauthoryear{{Pieri} et~al.,}{{Pieri}
  et~al.}{2014b}]{Pieri2014}
{Pieri} M.~M.,  et~al., 2014b, \mn@doi [\mnras] {10.1093/mnras/stu577}, \href
  {http://adsabs.harvard.edu/abs/2014MNRAS.441.1718P} {441, 1718}

\bibitem[\protect\citeauthoryear{{Pieri} et~al.,}{{Pieri}
  et~al.}{2014a}]{Pieri_etal2014}
{Pieri} M.~M.,  et~al., 2014a, \mn@doi [\mnras] {10.1093/mnras/stu577}, \href
  {https://ui-adsabs-harvard-edu.insu.bib.cnrs.fr/#abs/2014MNRAS.441.1718P}
  {441, 1718}

\bibitem[\protect\citeauthoryear{{Pieri} et~al.,}{{Pieri}
  et~al.}{2016}]{Pieri2016}
{Pieri} M.~M.,  et~al., 2016, in {Reyl{\'e}} C.,  {Richard} J.,  {Cambr{\'e}sy}
  L.,  {Deleuil} M.,  {P{\'e}contal} E.,  {Tresse} L.,   {Vauglin} I.,  eds,
  SF2A-2016: Proceedings of the Annual meeting of the French Society of
  Astronomy and Astrophysics. pp 259--266 (\mn@eprint {arXiv} {1611.09388})

\bibitem[\protect\citeauthoryear{{Planck Collaboration} et~al.,}{{Planck
  Collaboration} et~al.}{2016a}]{Planck2015}
{Planck Collaboration} et~al., 2016a, \mn@doi [\aap]
  {10.1051/0004-6361/201525830}, \href
  {http://adsabs.harvard.edu/abs/2016A%26A...594A..13P} {594, A13}

\bibitem[\protect\citeauthoryear{{Planck Collaboration} et~al.,}{{Planck
  Collaboration} et~al.}{2016b}]{Planck2016}
{Planck Collaboration} et~al., 2016b, \aap, 596, A108

\bibitem[\protect\citeauthoryear{{Rudie}, {Steidel}, {Shapley}  \&
  {Pettini}}{{Rudie} et~al.}{2013}]{Rudie2013}
{Rudie} G.~C.,  {Steidel} C.~C.,  {Shapley} A.~E.,   {Pettini} M.,  2013,
  \mn@doi [\apj] {10.1088/0004-637X/769/2/146}, \href
  {http://adsabs.harvard.edu/abs/2013ApJ...769..146R} {769, 146}

\bibitem[\protect\citeauthoryear{{Schaye}, {Aguirre}, {Kim}, {Theuns}, {Rauch}
  \& {Sargent}}{{Schaye} et~al.}{2003}]{Schaye2003}
{Schaye} J.,  {Aguirre} A.,  {Kim} T.-S.,  {Theuns} T.,  {Rauch} M.,
  {Sargent} W. L.~W.,  2003, \mn@doi [\apj] {10.1086/378044}, \href
  {https://ui-adsabs-harvard-edu.insu.bib.cnrs.fr/#abs/2003ApJ...596..768S}
  {596, 768}

\bibitem[\protect\citeauthoryear{{Segers}, {Oppenheimer}, {Schaye}  \&
  {Richings}}{{Segers} et~al.}{2017}]{Segers2017}
{Segers} M.~C.,  {Oppenheimer} B.~D.,  {Schaye} J.,   {Richings} A.~J.,  2017,
  \mn@doi [\mnras] {10.1093/mnras/stx1633}, \href
  {http://adsabs.harvard.edu/abs/2017MNRAS.471.1026S} {471, 1026}

\bibitem[\protect\citeauthoryear{{Shull}, {Tumlinson}, {Giroux}, {Kriss}  \&
  {Reimers}}{{Shull} et~al.}{2004}]{shull2004}
{Shull} J.~M.,  {Tumlinson} J.,  {Giroux} M.~L.,  {Kriss} G.~A.,   {Reimers}
  D.,  2004, \mn@doi [\apj] {10.1086/379924}, \href
  {http://adsabs.harvard.edu/abs/2004ApJ...600..570S} {600, 570}

\bibitem[\protect\citeauthoryear{{Shull}, {France}, {Danforth}, {Smith}  \&
  {Tumlinson}}{{Shull} et~al.}{2010}]{shull2010}
{Shull} J.~M.,  {France} K.,  {Danforth} C.~W.,  {Smith} B.,   {Tumlinson} J.,
  2010, \mn@doi [\apj] {10.1088/0004-637X/722/2/1312}, \href
  {http://adsabs.harvard.edu/abs/2010ApJ...722.1312S} {722, 1312}

\bibitem[\protect\citeauthoryear{{Songaila}, {Hu}  \& {Cowie}}{{Songaila}
  et~al.}{1995}]{Songaila1995}
{Songaila} A.,  {Hu} E.~M.,   {Cowie} L.~L.,  1995, \mn@doi [\nat]
  {10.1038/375124a0}, \href
  {https://ui-adsabs-harvard-edu.insu.bib.cnrs.fr/#abs/1995Natur.375..124S}
  {375, 124}

\bibitem[\protect\citeauthoryear{{Syphers} \& {Shull}}{{Syphers} \&
  {Shull}}{2013}]{Syphers2013}
{Syphers} D.,  {Shull} J.~M.,  2013, \mn@doi [\apj]
  {10.1088/0004-637X/765/2/119}, \href
  {http://adsabs.harvard.edu/abs/2013ApJ...765..119S} {765, 119}

\bibitem[\protect\citeauthoryear{{Syphers} \& {Shull}}{{Syphers} \&
  {Shull}}{2014}]{Syphers2014}
{Syphers} D.,  {Shull} J.~M.,  2014, \mn@doi [\apj]
  {10.1088/0004-637X/784/1/42}, \href
  {http://adsabs.harvard.edu/abs/2014ApJ...784...42S} {784, 42}

\bibitem[\protect\citeauthoryear{Syphers et~al.,}{Syphers
  et~al.}{2011}]{syphers2011}
Syphers D.,  et~al., 2011, \apj, 742, 99

\bibitem[\protect\citeauthoryear{{Weinberg}, {Katz}  \& {Hernquist}}{{Weinberg}
  et~al.}{1998}]{Weinberg1998}
{Weinberg} D.~H.,  {Katz} N.,   {Hernquist} L.,  1998, in Origins, ASP
  Conference Series, Vol. 148, 1998, ed. Charles E. Woodward, J. Michael Shull,
  and Harley A. Thronson, Jr. (1998), p.21. p.~21

\bibitem[\protect\citeauthoryear{{Worseck} \& {Wisotzki}}{{Worseck} \&
  {Wisotzki}}{2006}]{Worseck2006}
{Worseck} G.,  {Wisotzki} L.,  2006, \mn@doi [\aap]
  {10.1051/0004-6361:20054549}, \href
  {http://adsabs.harvard.edu/abs/2006A%26A...450..495W} {450, 495}

\bibitem[\protect\citeauthoryear{{Worseck}, {Fechner}, {Wisotzki}  \&
  {Dall'Aglio}}{{Worseck} et~al.}{2007}]{Worseck2007}
{Worseck} G.,  {Fechner} C.,  {Wisotzki} L.,   {Dall'Aglio} A.,  2007, \mn@doi
  [\aap] {10.1051/0004-6361:20077585}, \href
  {http://adsabs.harvard.edu/abs/2007A%26A...473..805W} {473, 805}

\bibitem[\protect\citeauthoryear{Worseck, Prochaska, Hennawi  \&
  McQuinn}{Worseck et~al.}{2016}]{Worseck2016}
Worseck G.,  Prochaska J.~X.,  Hennawi J.~F.,   McQuinn M.,  2016, \apj, 825,
  144

\bibitem[\protect\citeauthoryear{Zheng et~al.,}{Zheng et~al.}{2004}]{Zheng2004}
Zheng W.,  et~al., 2004, \apj, 605, 631

\makeatother
\end{thebibliography}

\bsp	
\label{lastpage}
\end{document}